\documentclass[traditabstract,doublecolumn]{aa}
\usepackage{graphicx}
\usepackage{txfonts}
\usepackage{savesym}
\savesymbol{rotate}
\usepackage{deluxetable} 
\restoresymbol{DTBL}{rotate}

\usepackage[english]{babel}
\usepackage{times,epsfig}
\usepackage{longtable}
\usepackage{url}
\usepackage{mathrsfs}
\usepackage{url}
\usepackage{hyperref}
\hypersetup{
    colorlinks = true,
    linkcolor = blue,
    anchorcolor = blue,
    citecolor = blue,
    filecolor = blue,
    urlcolor = blue
    }
    
\usepackage{ulem}  
\usepackage{placeins}

\usepackage[T1]{fontenc}
\usepackage[final]{microtype}

\newcommand{\orcid}[1]{\href{https://orcid.org/#1}{\includegraphics[width=10pt]{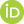}}}

\usepackage{float}
\usepackage{afterpage}
\bibpunct{(}{)}{;}{a}{}{,}

\newcommand{\nosne}{35}
\newcommand{\numberspectra}{170}

\begin{document} 
\authorrunning{Stritzinger, Holmbo, Morrell, et al.}
\titlerunning{CSP-I Spectroscopy of SE SNe.}

   \title{The Carnegie Supernova Project I. Optical spectroscopy of stripped-envelope supernovae}

\author{
  M. D. Stritzinger\inst{1}\orcid{0000-0002-5571-1833}
  \and
  S. Holmbo\inst{1}\orcid{0000-0002-3415-322X}
  \and
  N. Morrell\inst{2}\orcid{0000-0003-2535-3091}
  \and
   M. M. Phillips\inst{2}\orcid{0000-0003-2734-0796}
   \and
   C. R. Burns\inst{3}\orcid{0000-0003-4625-6629}
   \and 
    S. Castell\'{o}n\inst{2}
   \and 
   G. Folatelli\inst{4,5,6}\orcid{0000-0001-5247-1486}
   \and 
   M. Hamuy\inst{7,8}\orcid{0000-0001-7981-8320}
   \and 
     G. Leloudas\inst{9}\orcid{0000-0002-8597-0756}
     \and
   N.~B. Suntzeff\inst{10}\orcid{0000-0002-8102-181X}
  \and
  J. P. Anderson\inst{11}\orcid{0000-0003-0227-3451}
  \and 
   C. Ashall\inst{12}\orcid{0000-0002-5221-7557} 
  \and
   E. Baron \inst{13,1}\orcid{0000-0001-5393-1608}
   \and
   S. Boissier\inst{14}\orcid{0000-0002-9091-2366}
   \and 
   E. Y. Hsiao\inst{15}\orcid{0000-0003-1039-2928}
  \and
  E. Karamehmetoglu\inst{1}\orcid{0000-0001-6209-838X}
  \and
   F. Olivares\inst{16}\orcid{0000-0002-5115-6377}
}

\institute{
  Department of Physics and Astronomy, Aarhus University, Ny Munkegade 120, DK-8000 Aarhus C, Denmark\\ (\email{max@phys.au.dk})
  \and
  Carnegie Observatories, Las Campanas Observatory, Casilla 601, La Serena, Chile
  \and 
  Observatories of the Carnegie Institution for Science, 813 Santa Barbara St., Pasadena, CA 91101, USA
  \and 
  Facultad de Ciencias Astron\'omicas y Geof\'isicas, Universidad Nacional de La Plata, Paseo del Bosque S/N, B1900FWA, La Plata, Argentina
  \and 
  Instituto de Astrofísica de La Plata (IALP), CCT-CONICET-UNLP. Paseo del Bosque S/N, B1900FWA, La Plata, Argentina
  \and
  Kavli Institute for the Physics and Mathematics of the Universe(WPI), The University of Tokyo, 5-1-5 Kashiwanoha, Kashiwa, Chiba 277-8583, Japan
  \and 
  Fundaci\'{o}n Chilena de Astronomía, El Vergel 2252 \#1501, Santiago, Chile
\and
Hagler Institute for Advanced Studies, Texas A\&M University, College Station, TX 77843, USA
   \and
   DTU Space, National Space Institute, Technical University of Denmark, Elektrovej 327, 2800 Kgs. Lyngby, Denmark
  \and 
   George P. and Cynthia Woods Mitchell Institute for Fundamental Physics and Astronomy, Department of Physics and Astronomy, Texas A\&M University, College Station, TX 77843, USA
   \and
   European Southern Observatory, Alonso de C\'{o}rdova 3107, Casilla 19, Santiago, Chile
  \and 
  Institute for Astronomy, University of Hawai'i, 2680 Woodlawn Drive, Honolulu, HI 96822, USA  
  \and   
  Department of Physics and Astronomy, University of Oklahoma, 440  W. Brooks, Rm 100, Norman, OK 73019
  \and
  Aix Marseille Univ, CNRS, CNES, LAM, Marseille, France
  \and 
 Department of Physics, Florida State University, 77 Chieftain Way, Tallahassee, FL, 32306, USA
 \and 
 Instituto de Astronomia y Ciencias Planetarias, Universidad de Atacama, Copayapu 485, Copiapo, Chile
}
  \date{Received: February 20, 2022; Accepted: February 27, 2023}
 
\abstract{ 
  We present \numberspectra\  optical spectra of \nosne\ low-redshift stripped-envelope core-collapse supernovae observed by the \textit{Carnegie Supernova Project-I} between 2004 and 2009. The data extend from as early as $-$19 days (d) prior to the epoch of $B$-band maximum to  $+$322~d, with   the vast majority  obtained  during the so-called photospheric phase covering the weeks around peak luminosity. In addition to  histogram plots characterizing the redshift distribution, number of spectra per object, and the phase distribution of the sample,  spectroscopic classification is also provided following standard criteria. The CSP-I spectra are electronically available and a detailed analysis of the  data set is presented in a companion paper being the fifth and final paper of the series.} 
  
   \maketitle

\section{Introduction}

This is the fourth  in a series of five papers focusing on the presentation and analysis of a sample of stripped-envelope core-collapse supernovae (SE SNe) observed by the Carnegie Supernova Project~I (hereafter CSP-I; \citealt{2006PASP..118....2H}). 
SE SNe are the deaths  of massive stars that have undergone   prodigious mass loss over their evolutionary lifetimes.
Details associated with the physics of mass loss  and the explosion processes drives the  diversity characterizing the known spectroscopic SE SN subtypes. 
 In Paper~1 of the series  a detailed discussion on the contemporary understanding of SE SNe is presented   along with the broadband optical and near-infrared (NIR) photometry of the sample \citep[see][]{Stritz18}. 
 These data served as the basis in Paper~2 to improve upon our ability to estimate  host-reddening parameters \citep[see][]{stritzinger2018b}, while the results of both works were used in Paper~3 to characterize the photometric properties of the CSP-I sample  and estimate key explosion parameters via model comparisons \citep[see][]{taddia2018}. 

SE SNe are classified into spectroscopic subtypes based on the presence and/or absence of hydrogen (\ion{H}{i}) and helium (\ion{He}{i}) spectral features in  their optical spectra taken near the epoch of peak brightness \citep[see][for useful reviews]{filippenko1997,galyam2017}. 
Within  this framework  a spectroscopic  classification sequence emerges following the nomenclature: Type~IIb$\rightarrow$Type~Ib$\rightarrow$Type~Ic; hereafter SNe~IIb, Ib, and Ic. 
The progression of this sequence  reflects a higher degree of  mass stripping experienced by the  progenitor stars \citep{filippenko1997,Langer2012,Yoon2015,galyam2017,prentice2017,Shivvers2019}, though other factors  such as the prevalence of  $^{56}$Ni mixing in the SN ejecta and the explosion energy do contribute (see Paper~3, and below). 
As we briefly summarize in the following, significant diversity  exists amongst the observational properties of  various SE SN subtypes, prompting some authors to  suggest various extensions to the classical classification system  \citep[e.g.,][]{folatelli2014,prentice2017,Williamson2019}.  In this work, we adopt the traditional nomenclature SNe~IIb, Ib, Ic, and Ic-broad-line (BL).

Turning first to SNe~IIb, their spectra in the first days after explosion are blue and featureless (see the case of SN~2009K below). However, around a week past explosion they typically exhibit a prevalent  H$\alpha$ $\lambda6563$  feature  in their optical spectra along with other (less prominent)  Balmer series features, very much akin to the spectra of similar epoch H-rich SNe~II.
The \ion{H}{i} features are associated with a thin shell of hydrogen material (on the order of $\gtrsim0.5$~M$_{\sun}$; \citealt{2006A&A...450..305E}) surrounding  the  progenitor stars.
The low-$M_H$ shell  differentiates  SNe~IIb from classical SNe~II, and  ultimately leads to the diminishing strength of the hydrogen features    (particularly H$\alpha$) within days to weeks of the explosion. 
As the brightness of SNe~IIb evolves through  maximum light, they undergo a dramatic transition in their spectral properties with the  disappearance of Balmer features  accompanied by the emergence of  prevalent \ion{He}{i} $\lambda\lambda$4472, 5876, 6678, 7065, and 7281 features.  
SN~1993J was the first object with a spectral sequence that revealed a  the metamorphosis from a SN~II-like spectrum to  a SN~Ib spectrum \citep{1993ApJ...415L.103F, 1993Natur.365..232S}.
Today, SN~1993J is the archetype SN~IIb \citep{2000AJ....120.1487M}, and its progenitor was likely  a massive red supergiant star within a binary system \citep[e.g.,][]{2004Natur.427..129M}. 

By definition SNe~Ib show no signatures of hydrogen in their early spectra, though they do  typically show conspicuous  \ion{He}{i} lines that reach  prevalence  within several weeks post maximum \citep{1987ApJ...317..355H, Matheson2001}. Similar to the other SE SN subtypes, the spectra of SNe~Ib also contain features from various intermediate-mass elements, including \ion{Ca}{ii} H\&K, \ion{Na}{i} $\lambda\lambda$5890, 5896 doublet (blended with \ion{He}{i} $\lambda$5876), possibly the \ion{Si}{ii}  $\lambda$6355 doublet  blended with  \ion{C}{ii} $\lambda$6580 (and/or with a multitude of \ion{Fe}{ii} lines), \ion{O}{i} $\lambda$7444,  and the \ion{Ca}{ii} NIR triplet. In addition, a variety of iron lines are located at the blue end of the optical spectrum, most notably  the \ion{Fe}{ii} multiplet 42, whose features are often used as a  proxy of the bulk ejecta velocity \citep{2002ApJ...566.1005B}. SN~1984L served as  an early prototypical example of a SN~Ib \citep{Wheeler1985,Harkness1987}, and today there is an assortment of data on dozens of objects in the literature. 
Interestingly, several detailed case studies combined with radiative transfer calculations suggested  some  SNe~Ib progenitors likely retain a  hydrogen shell with $M_H$ $\sim$ 10$^{-3}-0.2$  M$_{\sun}$ \citep[e.g.,][]{2002ApJ...566.1005B, 2010ApJ...718..957J,hachinger2012}.
Moreover, in addition to normal SNe~Ib,  a  small number of objects are also known to harbor weak helium lines that slowly  emerge  and  strengthen over time. Examples of such objects include SN~1999ex \citep{hamuy2002}, SN~2007Y \citep{stritzinger2018b}, SN~2005bf \citep{folatelli2006}, SN~2008ax \citep{2011ApJ...739...41C, 2011MNRAS.413.2140T} and SN~2010as \citep{folatelli2014}. These objects  have  been referred to as  intermediate and/or transitional SNe~Ib/c  \citep[e.g.,][]{hamuy2002,leloudas2011}, while \citet{folatelli2014} suggested  they should form their own spectroscopic subtype, referring to them as flat-velocity SNe~IIb.
 
Rounding out the end of the classical spectroscopic sequence of SE SNe are the SNe~Ic and SNe~Ic-BL. These designations  apply to a mixed bag of peculiar transients, with the SNe~Ic-BL being the subset of objects  that exhibit broad spectral features.  In general, optical spectra of SNe~Ic lack both hydrogen and helium features, while prevalent  features associated with ions of \ion{O}{i}, \ion{Ca}{ii}, \ion{Fe}{ii}, and notably \ion{Si}{ii}, typically dominate the photospheric phase spectrum.
The broad spectral features seen in SNe~Ic-BL are thought to be the consequence  of  significant line blending produced by ejecta traveling  at relatively high velocities (i.e., $\gtrsim$ 30,000 km~s$^{-1}$. 
SN~1998bw was the first SN~Ic-BL associated with the afterglow of  a long-duration gamma ray burst (GRB; \citealt{1998Natur.395..670G,iwamoto1998,patat2001}). 
In other cases, SNe~Ic-BL have been linked to less energetic X-ray flashes such as SN~2006aj and  XRF060218 \citep{2006Natur.442.1008C,2006ApJ...643L..99M,2006ApJ...645L..21M,2006A&A...454..503S}. Alternatively, some SNe~Ic-BL such as   SN~2002ap \citep[e.g.,][]{2002ApJ...572L..61M, 2002MNRAS.332L..73G} and  SN~2009bb \citep{pignata2011} lack an association with  high-energy emission, though  they are accompanied by  evidence of central engine activity \citep[e.g.,][]{2010Natur.463..513S,2017ApJ...837..128W}.

 It is a matter of open discussion whether or not SNe~Ic are  devoid of helium (or even hydrogen for that matter (see, e.g., \citealt{Matheson2001,2006PASP..118..791B,hachinger2012,Williamson2019}), as even a small amount of helium could be retained and remain transparent  due to its high ionization potential \citep[see][]{piro2014}. 
 Indeed, \citet{Shahbandeh2022} recently identified a non-negligible fraction of their SNe~Ic sample exhibit signatures of \ion{He}{i} $\lambda$20581.  In addition, observations of SN~2016coi/ASASSN-16fp provide one such example where \ion{He}{i} may account for some features \citep{yamanaka2017,prentice2018,terreran2019}, and  \citet{yamanaka2017} suggested its classification could be revised to SN~Ib-BL. Prior to the discovery of SN~2016coi, the presence of helium was also suggested  in the SN~Ic-BL objects SN~2009bb \citep{pignata2011} and SN~2012ap \citep{milisavljevic2015}. These objects therefore appear to be transitional objects between SNe~Ic-BL and SNe~Ib.

Signatures of circumstellar interaction (CSI) driven emission features produced by the interaction between expanding SN ejecta and circumstellar material (CSM) have also been observed in post-maximum spectra of some SE SNe ranging from weeks to months post explosion \citep[e.g.,][]{tartaglia2021}  to even over a decade \citep[e,g.,][]{mauerhan2018}. 
In the case of SN~2017dio, early phase spectra presented by \citet{kuncarayakti2018} exhibit prevalent Balmer emission features superposed on a SN~Ic spectrum. In other instances,  narrow helium lines (often with a P Cygni profile) are observed in the early spectra of stripped SNe known as SNe~Ibn \citep[e.g.][]{pastorello2007,foley2007,Pastorello2008,Hosseinzadeh2017,Karamehmetoglu2021}. Finally, narrow carbon features have also been  detected in a handful of the so-called SNe~Icn \citep[e.g.,][]{fraser2021,galyam2021}. 

The timescales at which CSI occurs are largely  a consequence of the distribution and location of the CSM relative to the expanding ejecta, while the ions associated with the emission are largely dependent on the chemical composition of the CSM. With robust constraints on the emergence and duration of   the CSI features we are able to  put tight constraints on the  mass-loss history of the progenitors during the pre-SN phase (months to centuries),  while objects exhibiting late-phase emission  enable one to place constraints over much longer  evolutionary timescales. In addition to emission lines, CSI  produces additional energy deposition. This has been  inferred through signatures of optical broadband excesses \citep[e.g.,][]{sollerman2020}, radio emission \citep{maeda2021}, and a combination of optical, radio and X-ray emission \citep{pooley2019}.

Finally, we note that with the advent of high-cadence surveys, a variety of   hydrogen-deficient  SNe that exhibit rapid light-curve evolution have been discovered.  Examples  studied to date  include: SN~2002bj \citep{poznanski2010}, SN~2005ek \citep{drout2013}, SN~2010X \citep{kasliwal2010},   iPTF 14gqr \citep{de2018},  SN~2019bkc \citep{chen2020-bkc,prentice2020}, and SN~2019ehk \citep{nakaoka2021}. Despite their similarities to some SE SNe, no  single explosion model can satisfactorily reproduce the main observational properties of these fast-evolving cosmic explosions \citep[see][for a discussion]{chen2020-bkc}. 
 
In the following we present \numberspectra\ optical spectra  of \nosne\ SE SNe observed by the CSP-I. The majority of these objects are rather normal (see Paper 3) and enable us to perform a robust analysis on SE SN spectra (see Paper 5; \citealt{Holmbo2023}). 
A key contribution of the CSP-I sample to the literature sample is the quality of the spectra, characterized by high signal-to-noise that often extends through 9000~\AA, enabling an examination of the prominent  \ion{Ca}{ii} NIR triplet. In Sect.~\ref{sec:observations} details concerning  the  facilities used by the CSP-I to collect the data and the reduction processes are presented. In Sect.~\ref{sec:results} the spectra are presented, and their spectroscopic  subtype are determined.  We conclude with a summary in Sect.~\ref{sec:summary}. 

To date, optical spectroscopic samples consisting of hundreds of SE SN spectra  have been presented by the Berkeley SN group \citep{Matheson2001,Shivvers2019},  the CfA SN group \citep{Modjaz2014, Liu2016}, and the (i)PTF collaboration \citep{Fremling2018}. Turning to longer wavelengths, we also note that our follow-up project, dubbed CSP-II, recently released 75 NIR spectra of 34 SE SNe observed between 2011 and 2015 \citep{Shahbandeh2022}. The light curves and optical spectroscopy of the CSP-II SE SN sample will be the subject of a future publication.

\section{Observations}\label{sec:observations}

The CSP-I carried out five, a nine-month-long  observing campaigns between 2004 and 2009 centered on the southern hemisphere summer. By completion of the CSP-I, optical/NIR photometry and optical spectroscopy were obtained for several hundred SNe of all types, including
34 SE SNe largely discovered by targeted transient surveys (see Paper~1 for details). Table~\ref{table1} contains a compilation of key information of each object as well as SN~2009dq which was not a part of the light curve sample presented in Paper~1. This includes: the Internationcal Astronomical Union (IAU) name, the coordinates, the redshift ($z$) of the host galaxy, the spectral classification, the number of spectra obtained for each object, the phase range of the spectra, and an estimate of the time of  $t(B)_{max}$. Given our constraints to follow relatively bright SNe (i.e., peak apparent $m_V \leq 18$ mag) the targets are located  in nearby  galaxies. As demonstrated in the top panel of Fig.~\ref{fig:obs_red_no}, 34 of the objects have heliocentric redshifts in the range $z = 0.0046-0.0492$, while the luminous SN~2009ca (peak $M_V \approx -20$ mag) has a redshift of $z = 0.0957$. Additional panels in Fig.~\ref{fig:obs_red_no} contain histograms of the number of spectra obtained  per SN (middle panel) and the distribution of temporal phases with respect to $t(B)_{max}$ of the first epoch of observation (bottom panel).  

The CSP-I spectroscopy sample of SE SNe was obtained with a handful of telescopes and instruments located in Chile.
A journal of spectroscopic observations is provided in Table~\ref{table2}. This includes a listing of the spectra, date of observations, the observational facilities, and  the key spectral parameters: restframe phase relative to the epoch of $B$-band maximum, wavelength range, resolution, exposure time, and airmass. The vast majority of observations were conducted with facilities  at the Las Campanas Observatory with the 2.5~m du Pont telescope and the 6.5~m Magellan Baade and Clay telescopes. Most of our spectra were taken with the du Pont telescope equipped with the Wide Field reimaging CCD (WFCCD) camera.  A few spectra  were also taken with the  Boller \& Chivens (B\&C) spectrograph. Spectroscopy performed with the much larger-aperture Magellan telescopes made use of the Low Dispersion Survey Spectrograph 3 (LDSS3) and the Inamori Magellan Areal Camera and Spectrograph (IMACS). CSP-I also used the European Southern Observatory (ESO)  3.5~m New Technology Telescope (NTT) equipped with the ESO Multi-Mode Instrument (EMMI) and the 3.6~m telescope equipped with the ESO Faint Object Spectrograph and Camera (EFOSC), both located at the La Silla Observatory. In addition, the CSP-I also obtained three SE SN spectra  with the Ritchey-Chr{\'e}tien (RC) Cassegrain spectrograph  attached the 1.5~m telescope at the Cerro Tololo Inter-American Observatory (CTIO),   and  a half dozen (mostly late phase) spectra with the Gemini Multi-Object Spectrograph (GMOS) attached to the 8.1~m Gemini-South telescope on Cerro Pachon.

The reduction procedures followed to obtain   one-dimensional, flux-calibrated spectra are described in detail by \cite{2006PASP..118....2H}. In short,  a two-dimensional spectral image is  bias- and flat-field-corrected whereupon  the  SN trace is  extracted.  A  wavelength calibration solution determined from arc lamp exposures is  then applied to the one-dimensional extracted spectrum  followed by (when possible) the division of a telluric spectrum. Next the wavelength-corrected (and telluric-corrected) one-dimensional spectrum is multiplied by a nightly  sensitivity function determined from  observations of  one or more flux standard stars.  If multiple science exposures are obtained for a given object on the same night,  the  one-dimensional spectra  are averaged. Finally, prevalent cosmic rays are removed.  

\section{Results}\label{sec:results}
\subsection{Optical spectroscopy}

The sample consists of \numberspectra\ optical spectra of the 34 objects presented in Paper~1, as well  as two  spectra of the  Type~IIb SN~2009dq. The data sample covers a range of phases, extending from as early as $-$19 rest-frame days(d) relative to $t(B)_{max}$  (SN~2009K) to $+$310~d (SN~2008aq).\footnote{Estimates of $t(B)_{max}$ are  presented in Paper~3. In the present paper, all temporal phases are given in rest-frame days (d) relative to $t(B)_{max}$.} The histograms plotted in the middle and bottom panels of Fig.~\ref{fig:obs_red_no} indicate that an average of two to three spectra were obtained per object and that for the majority of objects the first spectrum was obtained around the time of the $B$-band maximum. Four  objects in the sample (i.e., SN~2007Y, SN~2008aq, SN~2009K and SN~2009bb) have a least one late-phase spectrum taken after $+$200~d. The nebular spectra of SN~2008aq and SN~2009K are presented for the first time and those of SN~2007Y and SN~2009bb were previously presented by  \citet{stritzinger2009} and \citet{pignata2011}, respectively.
The final extracted, one-dimensional spectra, normalized to their mean flux values are plotted in  Fig.~\ref{fig:raw_spectra}.\footnote{The spectra can be downloaded electronically from the CSP Pasadena-based web page at \url{http://csp.obs.carnegiescience.edu/data/}, and the Weizmann Interactive Supernova Data Repository (WISeREP; \citealt{Ofer2012}) \url{https://www.wiserep.org//object/}.} 
For presentation purposes, each plotted spectrum was smoothed using a median filter with a width of 5 pixels, while in a handful of objects widths of 7 pixels (SNe~2007kj and 2009dp) and 9 pixels (SNe~2006bf and SN~2006ep) were adopted. 

To assess the general quality of the flux calibration of our SE SN spectroscopic sample, we compared the observed broadband colors of the SNe  to the observed colors obtained from synthetic photometry. 
The inferred synthetic  colors were computed  for the entire WFCCD subset of spectra using the CSP-I system response functions that  define the natural photometric system of the Swope telescope \citep[see][]{2011AJ....142..156S,Krisciunas2017}.

The results of this exercise are plotted in Fig.~\ref{fig:phot_vs_synphot}, which shows the comparisons between synthetic and observed  ($g-r$),  ($r-i$), and ($V-i$) colors in the top panel. In the bottom panel  the differences between the observed and synthetic  colors are  plotted versus the corresponding synthetic colors. Comparison of the observed and synthetic colors reveals a fairly good agreement between  the ($g-r$) colors, exhibiting average differences of $+0.015$ mag (with an associated root-mean-square  uncertainty rms $= 0.101$ mag). Good agreement is also found  between the observed and synthetic  ($V-r$) colors (not plotted in Fig.~\ref{fig:phot_vs_synphot}), which exhibit an average difference of $-0.017$ mag (rms of 0.059 mag).

Less agreement is found between the observed and synthetic  colors  that include the $i$ band. Specifically, the average  difference between the  observed versus synthetic ($V-i$) and ($r-i$)  colors amounts to $+$0.055 mag (rms $= 0.078$ mag)  and $+$0.072 mag (rms $= 0.073$ mag), respectively. The systematic offsets between the observed and synthetic ($V-i$) and ($r-i$) colors are  evident in the bottom panel of Fig.~\ref{fig:phot_vs_synphot}. As a first step towards identifying the culprit of these offsets we eliminated spectra that contained prevalent telluric  absorption. The curated sample provided improved  ($V-i$) and ($r-i$) color offsets that decreased to  $+$0.034 mag  (rms $= 0.064$ mag) and $+$0.055 mag (rms $= 0.072$ mag), respectively. Although the offsets decreased, they are not entirely brought down to the levels found in the other color combinations that do not include the $i$~band. 

We further investigated the issue and conducted an additional sequence of tests. First we  examined whether poorly subtracted telluric absorption features could be a problem. To this end,  observed and synthetic broadband colors were compared for the nearly dozen  different spectrophotometric flux standards observed over the course of the CSP-I with the du Pont telescope equipped with WFCCD and which have had their telluric features removed following our standard procedures as prescribed by \citet{1999PASP..111.1426B}.
This test revealed average synthetic versus broadband color differences as follows:  ($g-r$) $= +0.002$ mag (rms $= 0.007$ mag), ($V-i$) $= +0.002$ mag  (rms $= 0.028$ mag), and ($r-i$) $= +0.006$ mag (rms $= 0.024$ mag). 
The close agreement between these color indices reveals that, at least in the case of spectrophotometric standard star observations, our telluric removal technique produces robust results.

We also considered the accuracy of the CSP-I $i$-band system response function used to compute synthetic photometry as its measurement was not as robust as those of the other bands  \citep{Rheault2014}. Indeed, the $i$-band system response function was only scanned a single time and suffered illumination issues at red wavelengths due to the light source used. To this end, synthetic colors were computed with the CSP-I system response functions using   the atlas of spectrophotometric  Landolt standard stars \citep{2005PASP..117..810S}, which  have both \citet{1992AJ....104..340L} and \citet{2002AJ....123.2121S} broadband photometry. Comparison between the resulting synthetic colors to those inferred from the broadband photometry of these stars yields average color differences as follows: ($g-r$) $= -0.005$ mag  (rms $= 0.010$ mag), ($V-i$) $= -0.031$ mag  (rms $= 0.018$ mag), and ($r-i$) $= -0.006$ mag (rms $= 0.012$ mag).
 
 Although we are unable to conclusively identify the culprit driving the offsets with the $i$-band colors, we speculate it is linked to the accuracy of the $i$-band system response function. Indeed, as noted by \citet{Rheault2014}, the $i$-band system response function  proved to be more difficult to measure and less accurate than the other bands due to the low illumination of the light source used in the scanning process. In summary, we find good agreement between the observed and synthetic colors that do not include the $i$~band.

\subsection{Spectroscopic classification}

We now turn to the spectral classification of objects in the CSP-I SE SN sample. If  the spectrum of a given object lacked prevalent hydrogen and/or \ion{He}{i} lines, we turned to guidance from the spectral template comparison program SNID (SuperNova IDentification; \citealt{2007AIPC..924..312B}). In doing so, we made use of the expanded set of SE SN templates presented by  \citet{Liu2014}. 
The spectroscopic classification of the sample is reported in Table~\ref{table1} and reveals a breakdown between subtypes as follows: eleven SNe~IIb, twelve  SNe~Ib, ten SNe~Ic, and two SNe~Ic-BL. As a caveat to these  classifications, we note that traditionally it would not  be entirely possible to exclude a SN~IIb classification for the Type~Ib SNe~2004ew, 2006ep, and 2008gc as their earliest spectra are post-maximum (see Table~\ref{table1}). However, as demonstrated in Paper~5, from pseudo-equivalent width  measurements of certain spectral features and/or through the use of principal component analysis, these objects are confirmed to be SNe~Ib. 
Similarly the distinction between some SNe~Ib and SNe~Ic is not always clear cut as was found in the cases of   SNe~2005em and 2009dt, whose spectral phase coverage is  limited to only good spectra taken around or before maximum (see Table~\ref{table1}).  However, through the use of pEW measurements of \ion{O}{i} and/or PCA these two objects are found to be consistent with a SN~Ic classification. 

In assigning classifications to SNe~IIb no differentiation was made between objects thought to be characterized by extended or compact progenitors since such detailed subtyping is reliant on radio  emission limits and/or detections \citep[e.g.,][]{2010ApJ...711L..40C}. In the case of SN~2009bb, previous papers by our group and collaborators have noted  it exhibits  broad spectral features akin to other SNe~Ic-BL  \citep{pignata2011}  and that it displays signatures of a relativistic outflow possibly linked to a central engine \citep{2010Natur.463..513S}. 
SN~2009ca is also a noteworthy object as its peak luminosity reached a value of at least  $5\times10^{43}$ ergs s$^{-1}$, which is reminiscent of SN~1992ar \citep{Clocchiatti2000}. This value is  more than a factor of 10 larger compared to the average peak luminosity values determined from the literature sample of normal SNe~Ic  and a factor of 3 higher compared to the brightest of the gamma-ray bursts associated SNe~Ic-BL, such as SN~2012bz \citep{Schulze2014}. Indeed, SN~2009ca appears to be a super-luminous supernova \citep[see][]{taddia2018}. 

\section{Summary}
\label{sec:summary}

This paper presents \numberspectra\ optical spectra of \nosne\ SE SNe observed by CSP-I. The majority of the spectra were taken with facilities at the Las Campanas Observatory before or within a week after the epoch of $B$-band maximum. Nebular spectra are also presented for a  subset of five objects. In general the spectra are of high quality and often cover the spectral range to around 9000~\AA. In the companion paper presented by \citet[][i.e., Paper 5]{Holmbo2023} these data are used to construct  mean template spectral sequences for the IIb, Ib, and Ic SE SN subtypes, study key line diagnostics, and   perform a robust principle component analysis, leading to a fast and accurate method for classifying SE SN subtypes using only a  single spectrum taken at either early or post-maximum epochs.

\begin{acknowledgements}

The CSP has received support from the National Science Foundation (USA) under grants AST--0306969, AST--0607438, AST--1008343, AST--1613426, AST--1613455, and AST-1613472. 
The Aarhus supernova group is funding in part by a research Project 1 grant from the Independent Research Fund Denmark (IRFD grant numbers 8021-00170B and 10.46540/2032-00022B), and by a VILLUM FONDEN Experiment (grant number 28021). GL is supported by a Villum Young Investigator fellowship (grant number 19054) from the VILLUM FONDEN. This research has made use of the NASA/IPAC Extragalactic Database (NED), which is operated by the Jet Propulsion Laboratory, California Institute of Technology, under contract with the National Aeronautics and Space Administration. Based on observations collected  with the Magellan Clay and Baade 6.5-m telescopes and and the du Pont 2.5-m telescope located on Las Campanas Observatory, Chile;  the European Organization for Astronomical Research in the Southern Hemisphere (ESO), Chile, Programs 076.A-0156, 078.D-0048, 080.A-0516, 082.A-0526, and 380.D-0272; the 1.5 m telescope at CTIO, Chile; and the Gemini Observatory, Cerro Pachon, Chile (Gemini Programs GS-2008B-Q-8-77).

\end{acknowledgements}

\bibliographystyle{aa}
\bibliography{ms.bib}

\clearpage
\begin{deluxetable}{lccccccc}
\tabletypesize{\tiny}
\tablecolumns{8}
\tablewidth{0pt}
\tablecaption{Spectroscopic sample of CSP-I SE SNe.\label{table1}}
\tablehead{
\colhead{SN} &
\colhead{$\alpha (2000)$} &
\colhead{$\delta (2000)$} &
\colhead{Redshift$^{a}$} &
\colhead{Spectral} &
\colhead{No. of} &
\colhead{Phase} &
\colhead{$t(B)_{\rm max}$}\\
\colhead{} &
\colhead{} &
\colhead{} &
\colhead{$z$} &
\colhead{Type} &
\colhead{Spectra} &
\colhead{Range} &
\colhead{JD+2450000}}
\startdata
  2004ew & 02:05:06.17 & $-$55:06:31.6 & 0.0218 & Ib     &  4 & $+26\cdots +$81 &  $3276.11\pm2.03$ \\
  2004ex & 00:38:10.17 & $+$02:43:16.9 & 0.0176 & IIb    &  5 & $ -4\cdots +$41 &  $3306.69\pm0.03$ \\
  2004fe & 00:30:11.27 & $+$02:05:23.5 & 0.0179 & Ic     &  4 & $ +9\cdots +$35 &  $3316.79\pm0.11$ \\
  2004ff & 04:58:46.19 & $-$21:34:12.0 & 0.0227 & IIb    &  3 & $+14\cdots +$30 &  $3312.60\pm0.03$ \\
  2004gq & 05:12:04.81 & $-$15:40:54.2 & 0.0065 & Ib     &  3 & $ -5\cdots +$24 &  $3357.90\pm0.02$ \\
  2004gt & 12:01:50.37 & $-$18:52:12.7 & 0.0055 & Ic     &  4 & $ -2\cdots +$49 &  $3360.37\pm0.08$ \\
  2004gv & 02:13:37.42 & $-$00:43:05.8 & 0.0199 & Ib     &  1 & $ -7\cdots -$7  &  $3365.27\pm0.31$ \\
  2005Q  & 01:30:03.51 & $-$42:40:48.4 & 0.0224 & IIb    &  2 & $ -1\cdots +$7  &  $3406.10\pm0.04$ \\
  2005aw & 19:15:17.44 & $-$54:08:24.9 & 0.0095 & Ic     &  5 & $ +7\cdots +$23 &  $3456.48\pm1.37$ \\
  2005bf & 10:23:57.27 & $-$03:11:28.6 & 0.0189 & Ib     &  9 & $ -7\cdots +$52 &  $3474.79\pm0.28$ \\ 
  2005bj & 16:49:44.74 & $+$17:51:48.7 & 0.0222 & IIb    &  2 & $ +1\cdots +$4  &  $3471.89\pm1.37$ \\
  2005em & 03:13:47.71 & $-$00:14:37.0 & 0.0260 & Ic     &  1 & $ +0\cdots +$0  &  $3648.97\pm1.36$ \\
  2006T  & 09:54:30.21 & $-$25:42:29.3 & 0.0081 & IIb    & 11 & $ +0\cdots +$70 &  $3779.61\pm0.01$ \\
  2006ba & 09:43:13.40 & $-$09:36:53.0 & 0.0191 & IIb    &  3 & $ +4\cdots +$28 &  $3820.92\pm1.36$ \\
  2006bf & 12:58:50.68 & $+$09:39:30.1 & 0.0239 & IIb    &  5 & $ +7\cdots $+33 &  $3817.46\pm2.03$ \\
  2006ep & 00:41:24.88 & $+$25:29:46.7 & 0.0151 & Ib     &  3 & $+19\cdots +$34 &  $3985.41\pm0.05$ \\
  2006fo & 02:32:38.89 & $+$00:37:03.0 & 0.0207 & Ib     &  3 & $ +2\cdots +$18 &  $4003.17\pm1.36$ \\
  2006ir & 23:04:35.68 & $+$07:36:21.5 & 0.0200 & Ic     &  2 & $+19\cdots +$43 &  $3999.07\pm2.03$ \\
  2006lc & 22:44:24.48 & $-$00:09:53.5 & 0.0162 & Ib     &  1 & $ +1\cdots +$1  &  $4041.33\pm0.04$ \\
  2007C  & 13:08:49.30 & $-$06:47:01.0 & 0.0056 & Ib     &  9 & $ -1\cdots +$92 &  $4115.25\pm1.37$ \\
  2007Y  & 03:02:35.92 & $-$22:53:50.1 & 0.0046 & Ib     & 12 & $-12\cdots +$271&  $4162.99\pm0.01$ \\
  2007ag & 10:01:35.99 & $+$21:36:42.0 & 0.0207 & Ib     &  2 & $ +7\cdots +$10 &  $4163.29\pm1.42$ \\
  2007hn & 21:02:46.85 & $-$04:05:25.2 & 0.0273$^{\dag}$ & Ic     &  4 & $+10\cdots +$56 &  $4351.63\pm1.41$ \\
  2007kj & 00:01:19.58 & $+$13:06:30.6 & 0.0179 & Ib     &  5 & $ -4\cdots +$40 &  $4380.91\pm0.03$ \\
  2007rz & 04:31:10.84 & $+$07:37:51.5 & 0.0130 & Ic     &  2 & $ +7\cdots +$32 &  $4437.31\pm2.03$ \\
  2008aq & 12:50:30.42 & $-$10:52:01.4 & 0.0080 & IIb    & 12 & $ -5\cdots +$308&  $4531.15\pm0.02$ \\
  2008gc & 02:10:36.63 & $-$53:45:59.5 & 0.0492$^{\dag}$ & Ib     &  6 & $ +9\cdots +$117&  $4744.26\pm1.37$ \\
  2008hh & 01:26:03.65 & $+$11:26:26.5 & 0.0194 & Ic     &  3 & $ +2\cdots +$31 &  $4791.50\pm2.03$ \\
  2009K  & 04:36:36.77 & $-$00:08:35.6 & 0.0117 & IIb    &  8 & $-18\cdots +$282&  $4867.22\pm0.01$ \\
  2009Z  & 14:01:53.61 & $-$01:20:30.2 & 0.0248 & IIb    &  8 & $ -6\cdots +$41 &  $4876.94\pm0.01$ \\
  2009bb & 10:31:33.87 & $-$39:57:30.0 & 0.0099 & Ic-BL  & 18 & $ -1\cdots +$308&  $4920.05\pm0.02$ \\
  2009ca & 21:26:22.20 & $-$40:51:48.6 & 0.0957$^{\dag}$ & Ic-BL  &  3 & $ +3\cdots +$16 &  $4926.17\pm2.03$ \\
  2009dp & 20:26:52.69 & $-$18:37:04.2 & 0.0232 & Ic     &  3 & $ +2\cdots +$34 &  $4949.81\pm2.03$ \\
  2009dq & 10:08:49.94 & $-$67:01:57.3 & 0.0047$^{\dag}$ & IIb    &  2 & $ -9\cdots +$14 &  $4960.50\pm2.00$ \\ 
  2009dt & 22:10:09.27 & $-$36:05:42.6 & 0.0104 & Ic     &  2 & $ -6\cdots -$16 &  $4958.31\pm1.38$ \\
\enddata
\tablenotetext{a}{Heliocentric redshifts were retrieved from NED, or if not, then as determined from host-galaxy
emission lines in optical spectra. The latter of these are indicated by a \dag. }
\end{deluxetable}

\clearpage
\setcounter{figure}{0}
\begin{figure}
  \resizebox{\hsize}{!}
    {\includegraphics[]{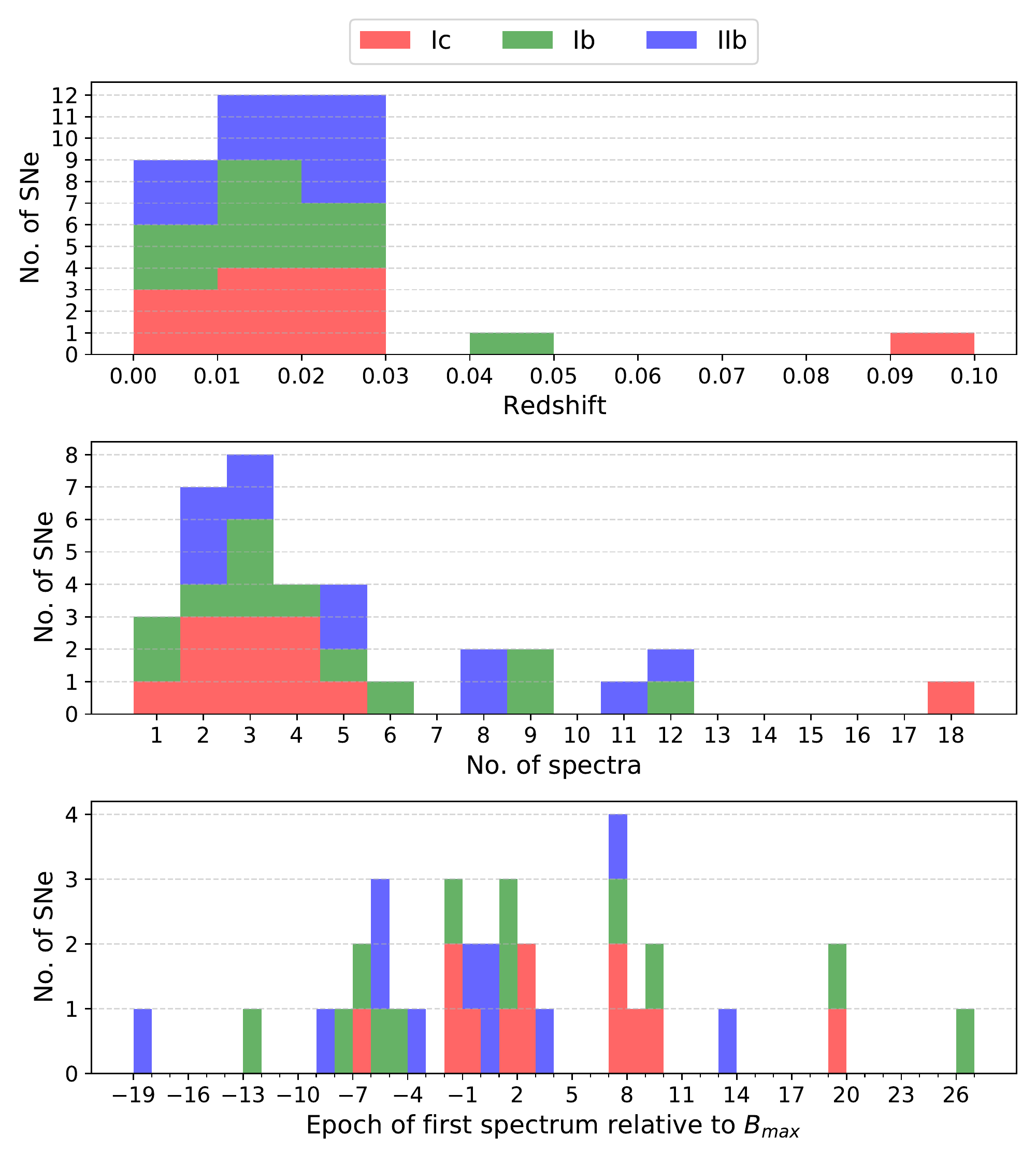}}
  \caption{Breakdown of the sample. \textit{Top panel:} Distribution of redshift ($z$) for the sample, as obtained from host-galaxy redshift measurements provided by the NASA/IPAC Extragalactic Database (NED). 
\textit{Middle panel:}  Distribution of the number of spectra per SN for the sample.
\textit{Bottom panel:} Temporal phase distribution of the first epoch of observation for the CSP-I SE SN spectroscopy sample.}
  \label{fig:obs_red_no}
\end{figure}

\clearpage
\setcounter{figure}{1}
\begin{figure*}
	\setlength\arraycolsep{0pt}
	\renewcommand{\arraystretch}{0}
	\centering$
  \begin{array}{cc}
    \includegraphics[width=.45\linewidth]{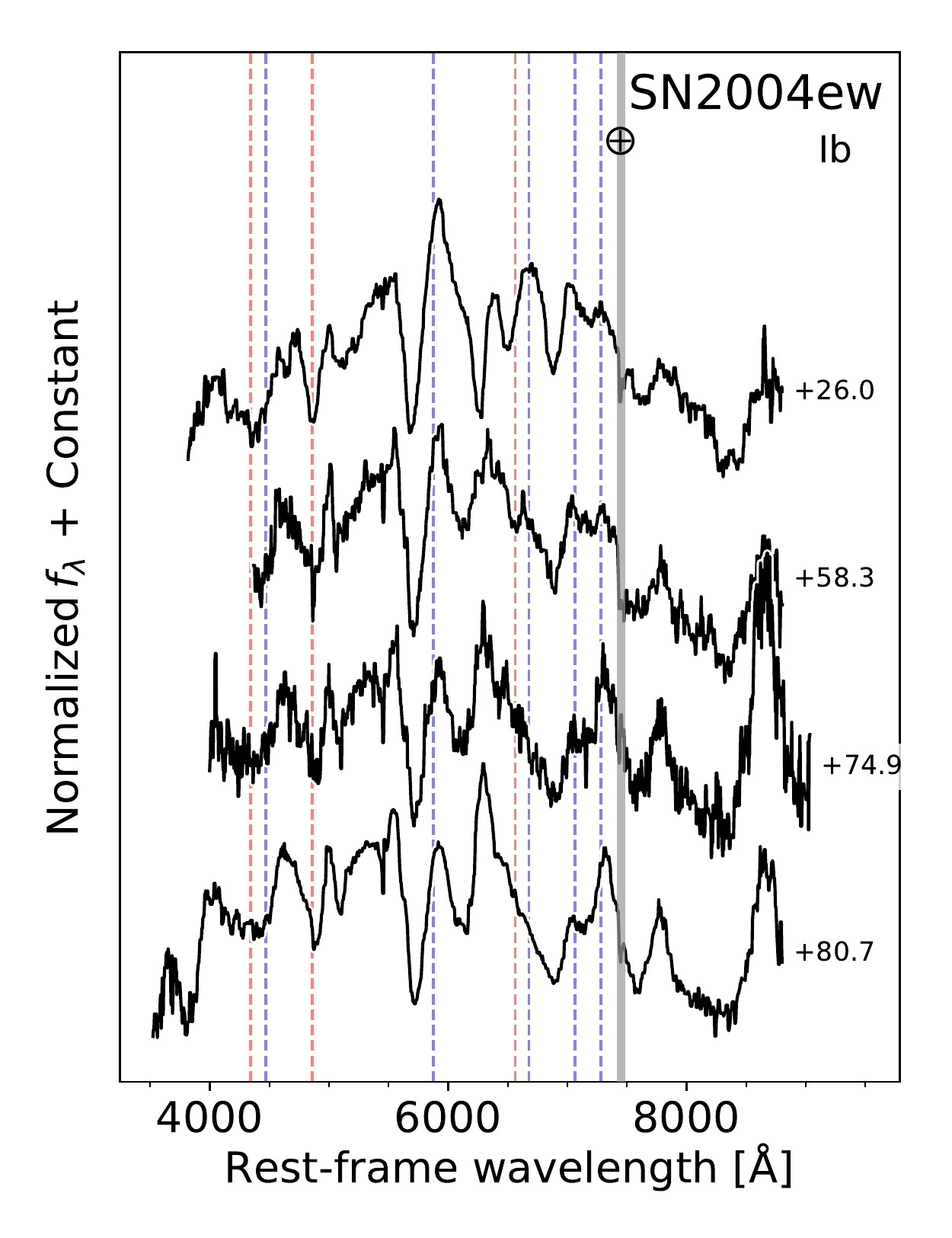} &
    \includegraphics[width=.45\linewidth]{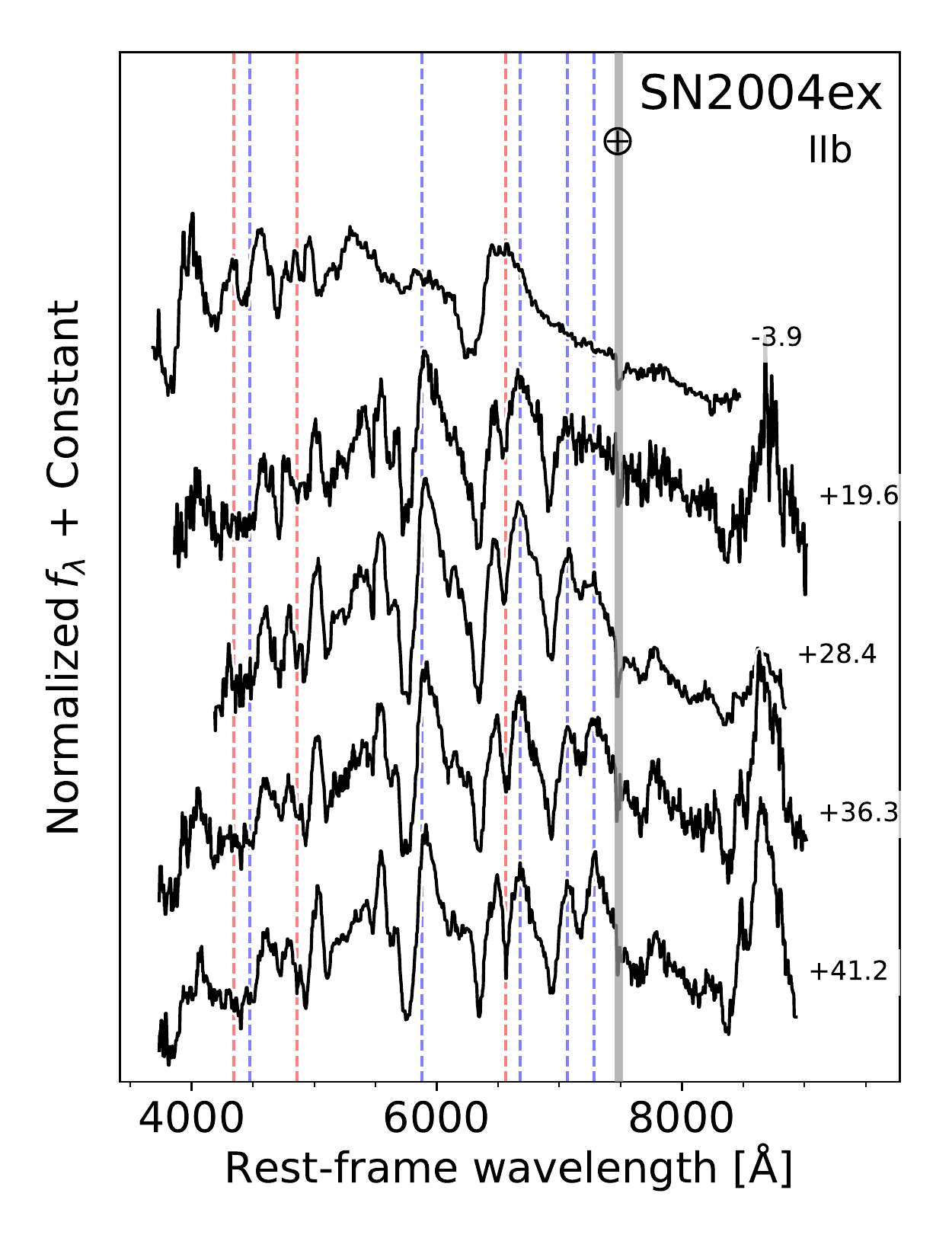} \\
    \includegraphics[width=.45\linewidth]{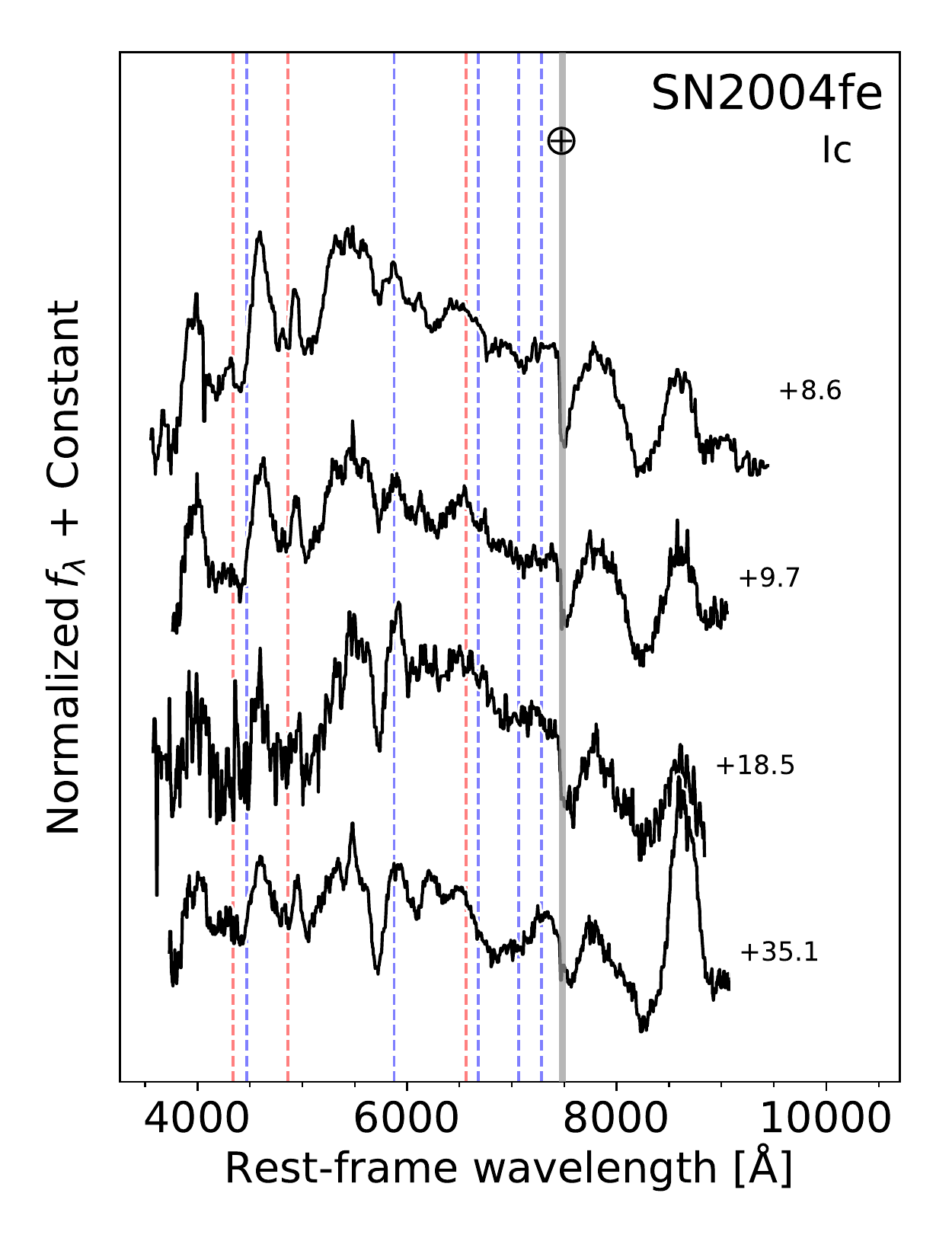} &
    \includegraphics[width=.45\linewidth]{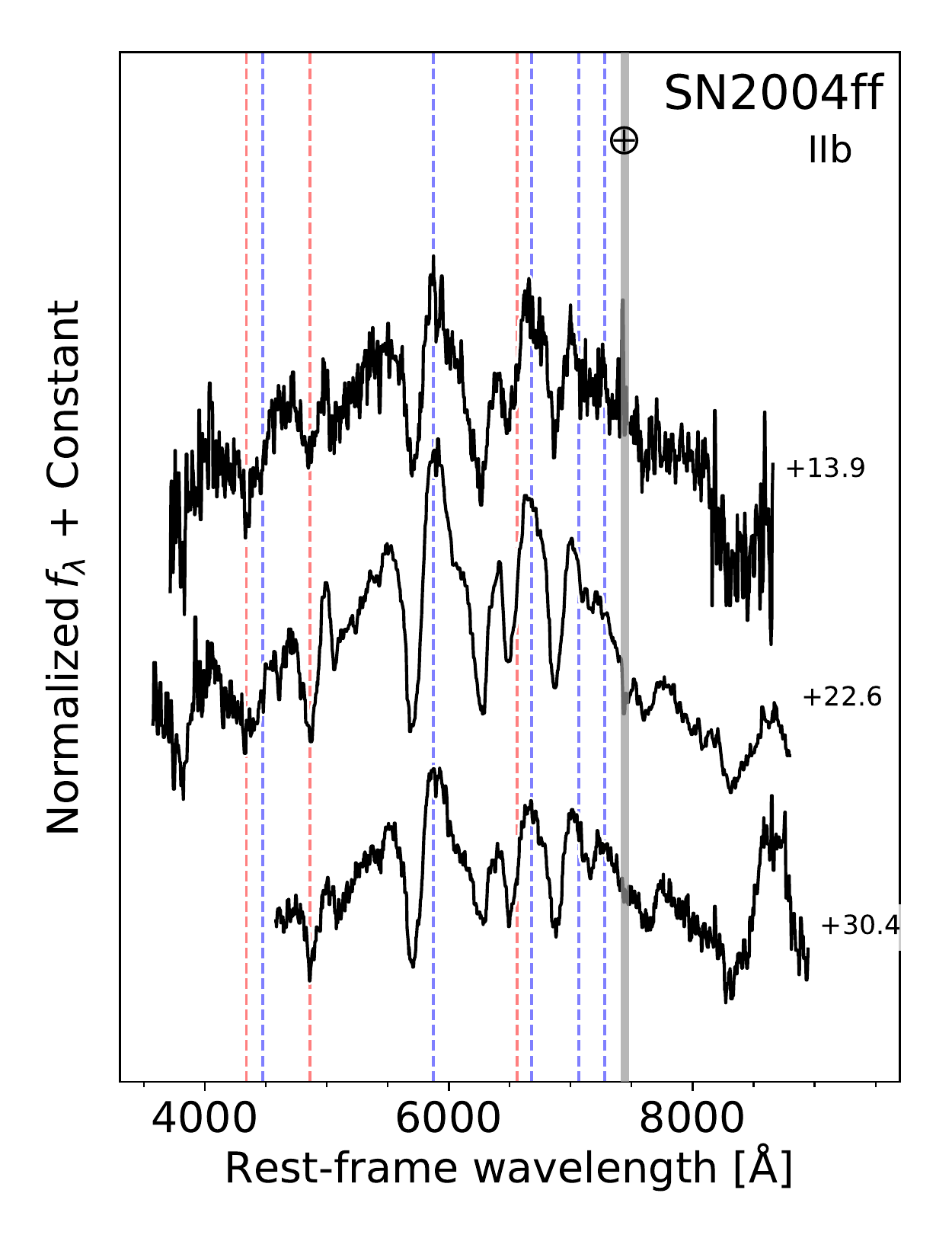} \\
  \end{array}$
  \caption{Optical spectroscopy of the CSP-I SE SN sample.  Each spectrum has been normalized to its mean flux values and plotted  in units of $f_\lambda$. For presentation purposes additive constants have also been applied. The phase of each spectrum relative to the epoch of the $B$-band maximum is provided next to each spectrum and the spectroscopic subtype is provided below the SN designation at the top right. Vertical dashed blue lines indicate rest wavelengths of  \ion{He}{i} $\lambda\lambda$4472, 5876, 6678, 7065, 7281 lines, vertical dashed red lines are H$\gamma$, H$\beta$, and H$\alpha$, and the vertical gray line indicates the location of a  prominent telluric feature.}
\end{figure*}

\setcounter{figure}{1}
\begin{figure*}
	\setlength\arraycolsep{0pt}
	\renewcommand{\arraystretch}{0}
	\centering$
  \begin{array}{cc}
    \includegraphics[width=.45\linewidth]{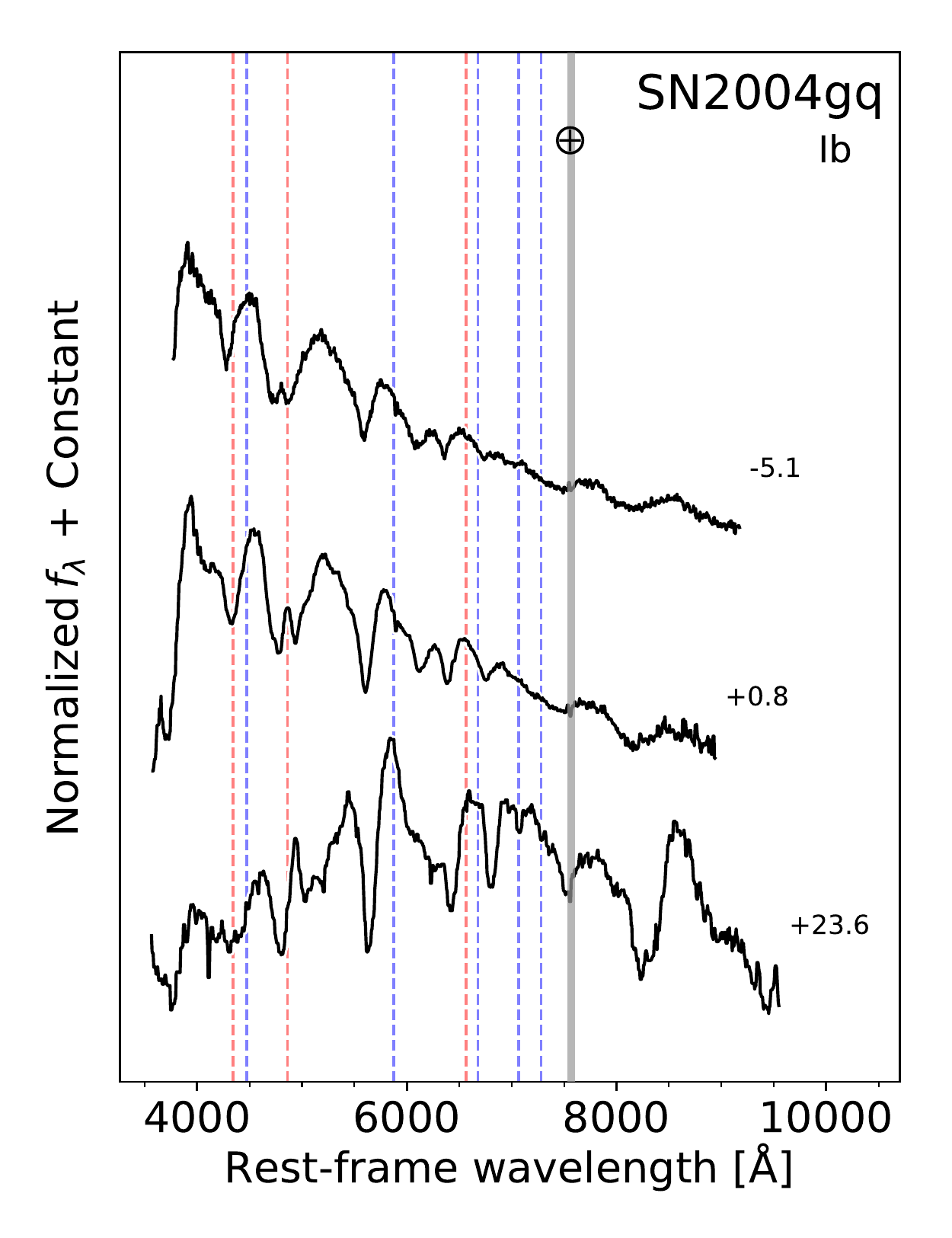} &
    \includegraphics[width=.45\linewidth]{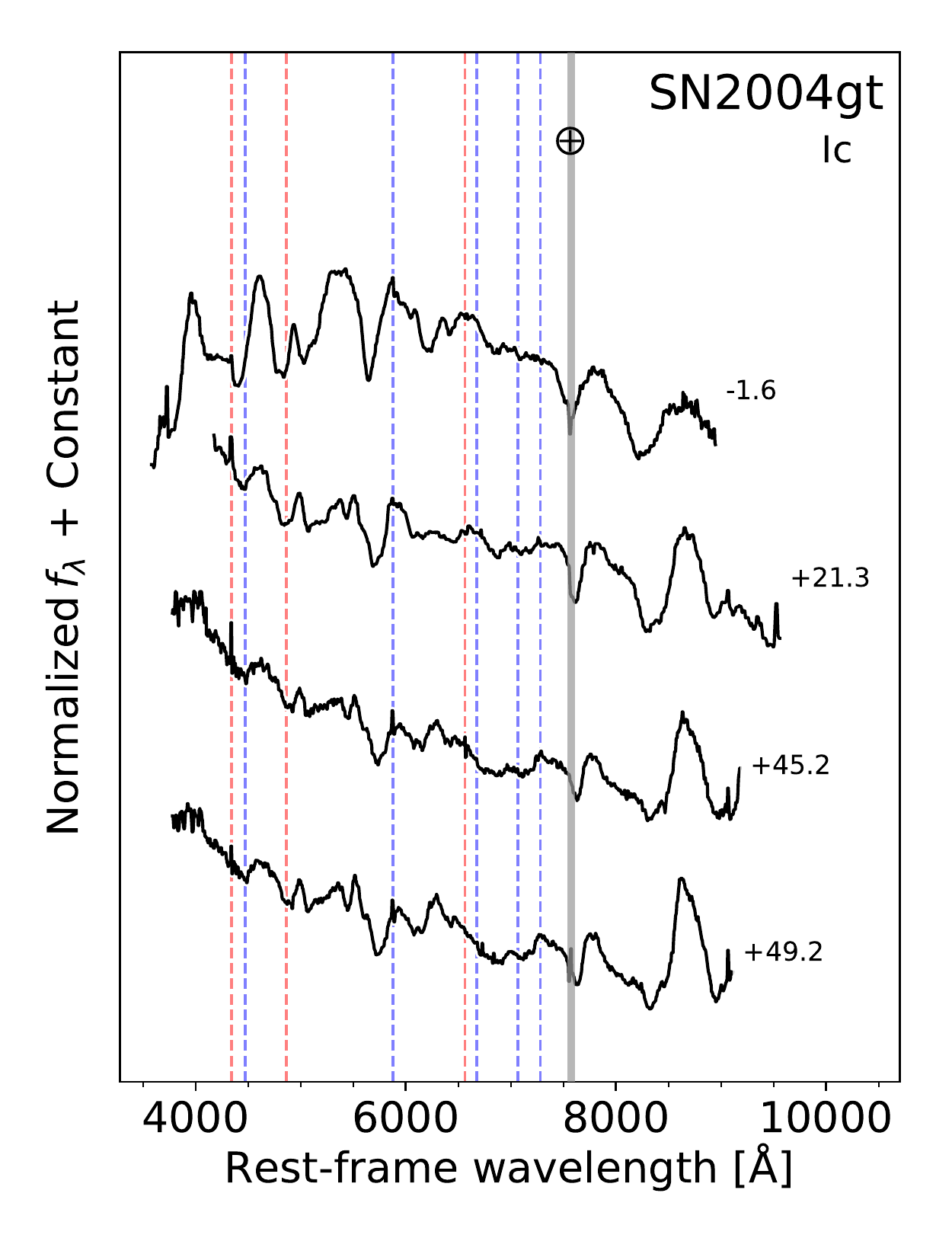} \\
    \includegraphics[width=.45\linewidth]{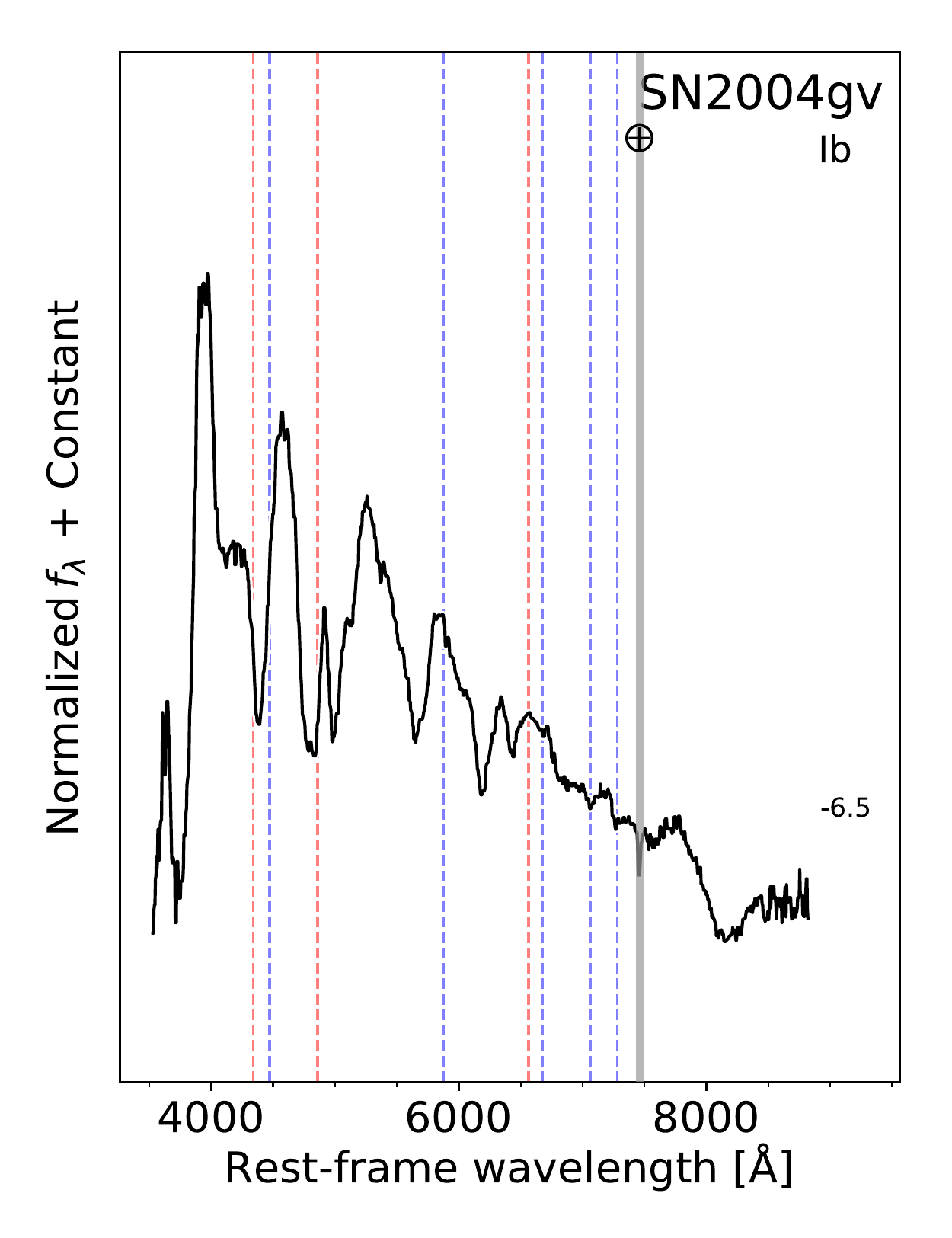} &
    \includegraphics[width=.45\linewidth]{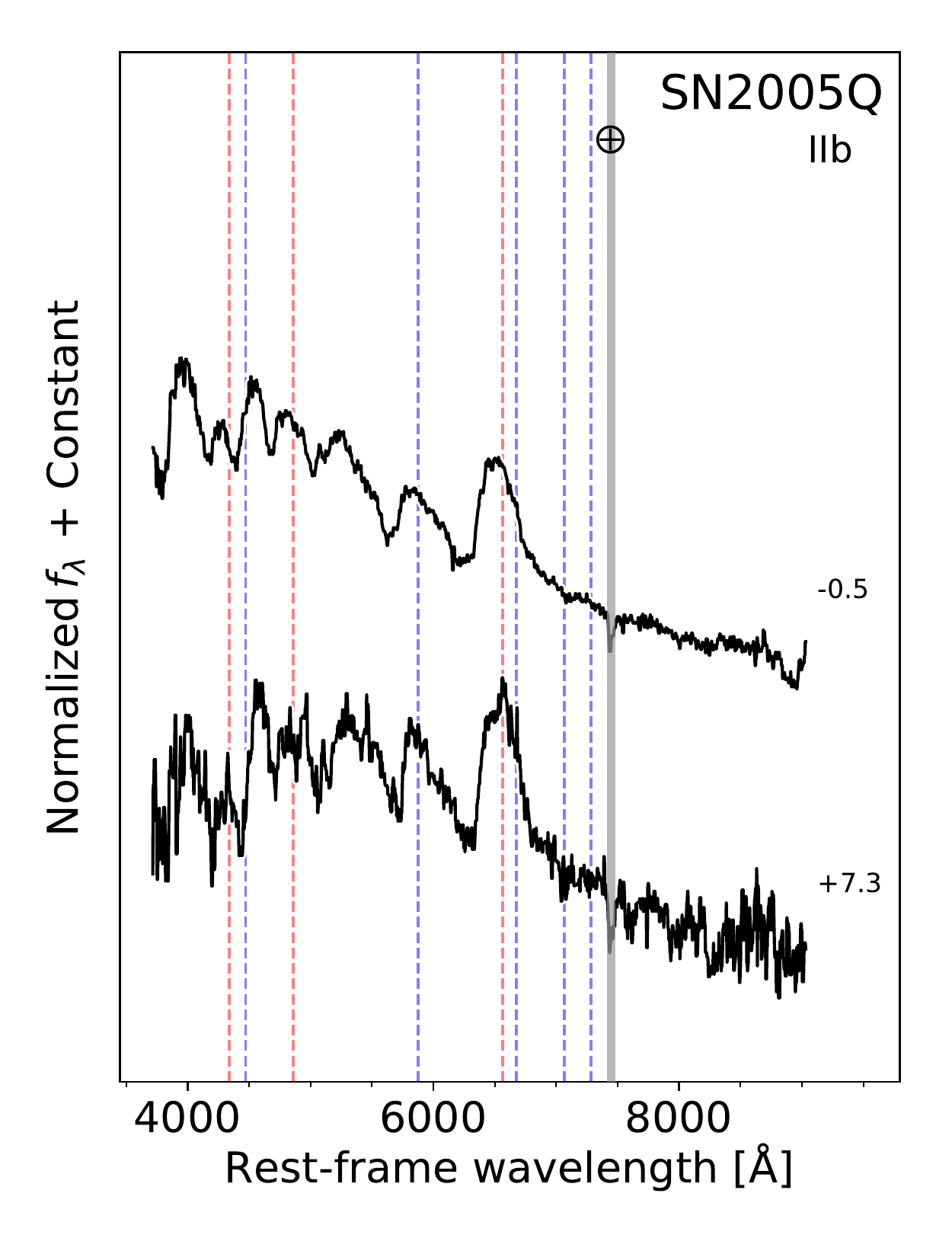} \\
  \end{array}$
   \caption{Continued.
}
\end{figure*}

\setcounter{figure}{1}
\begin{figure*}
	\setlength\arraycolsep{0pt}
	\renewcommand{\arraystretch}{0}
	\centering$
  \begin{array}{cc}
    \includegraphics[width=.45\linewidth]{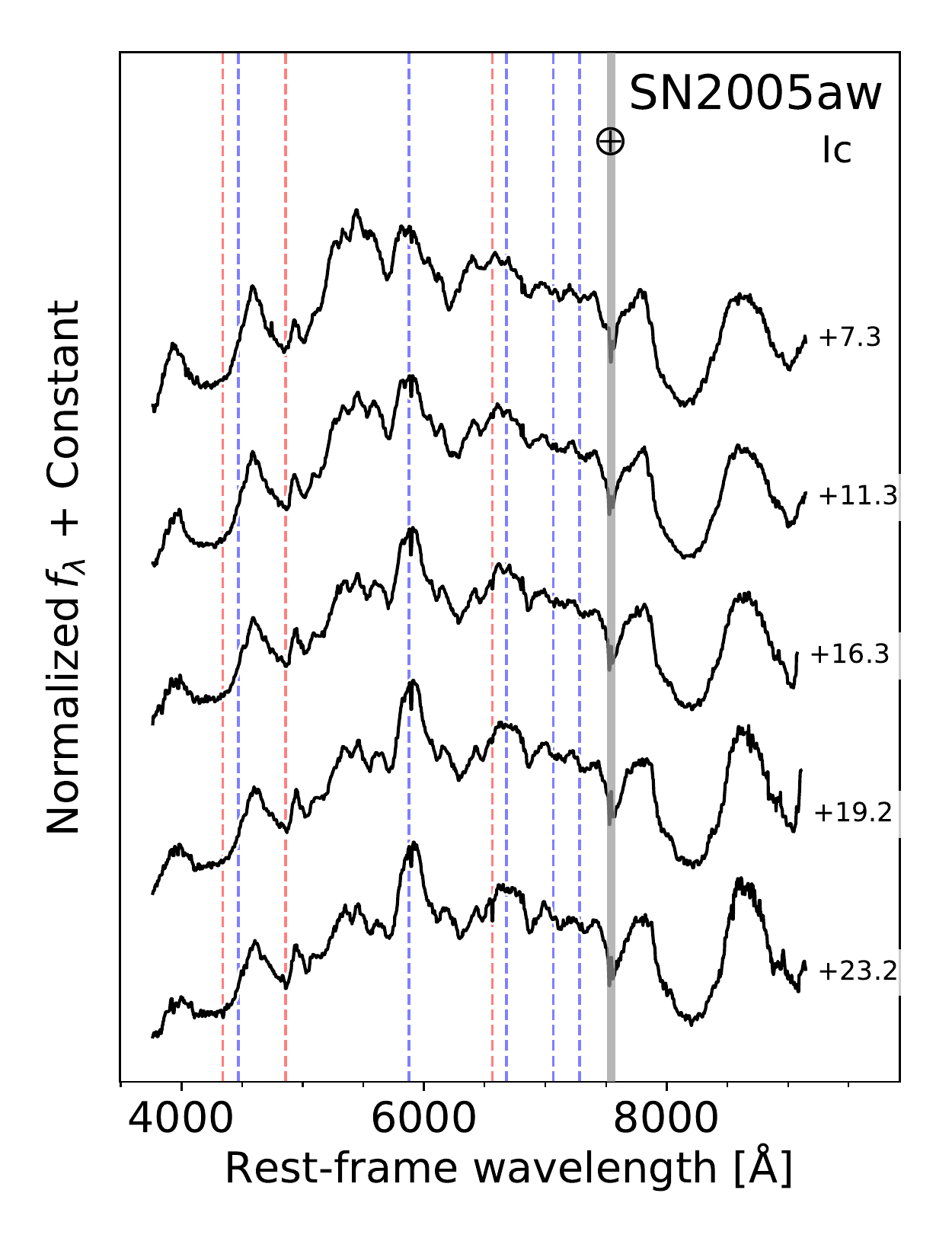} &
    \includegraphics[width=.45\linewidth]{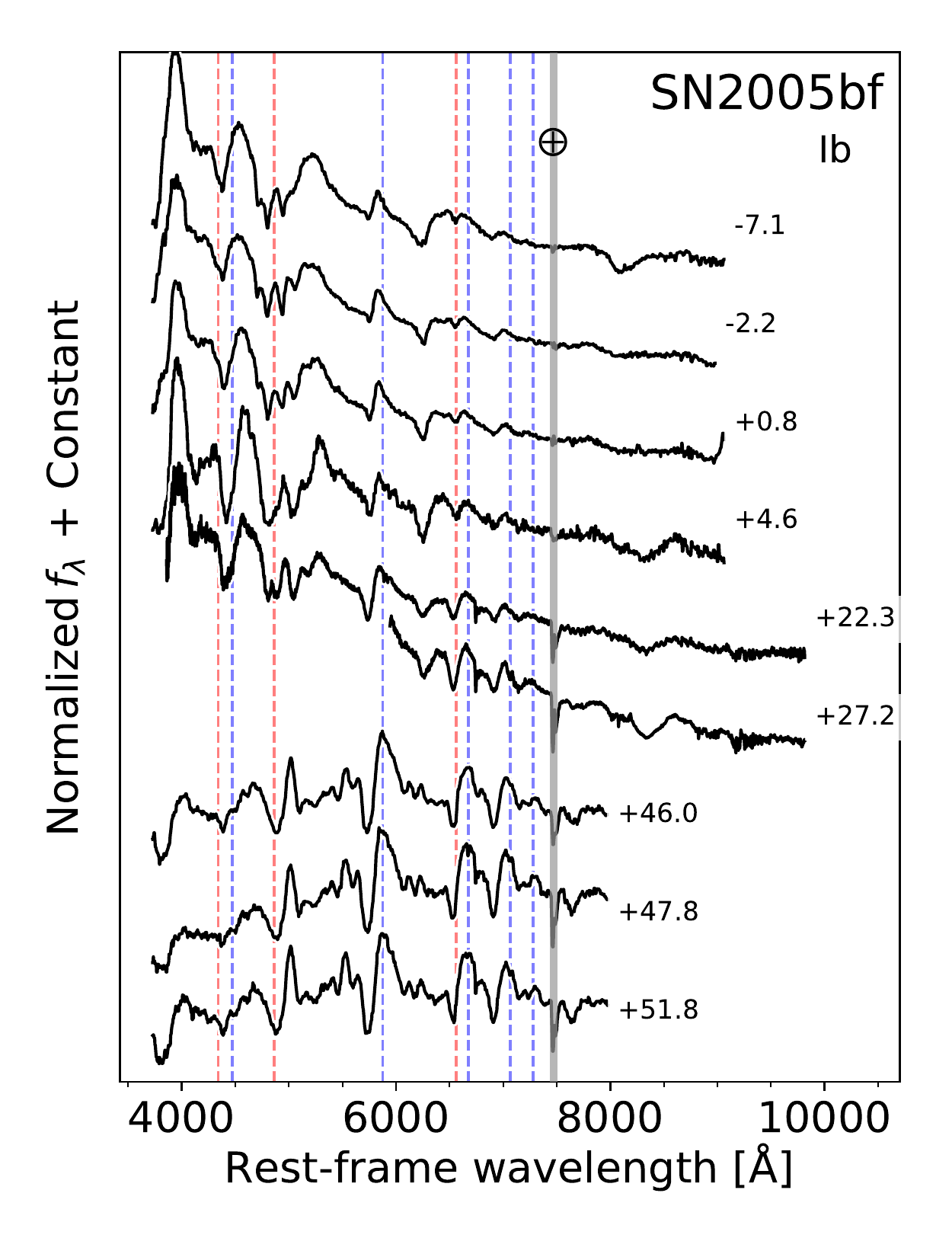} \\
    \includegraphics[width=.45\linewidth]{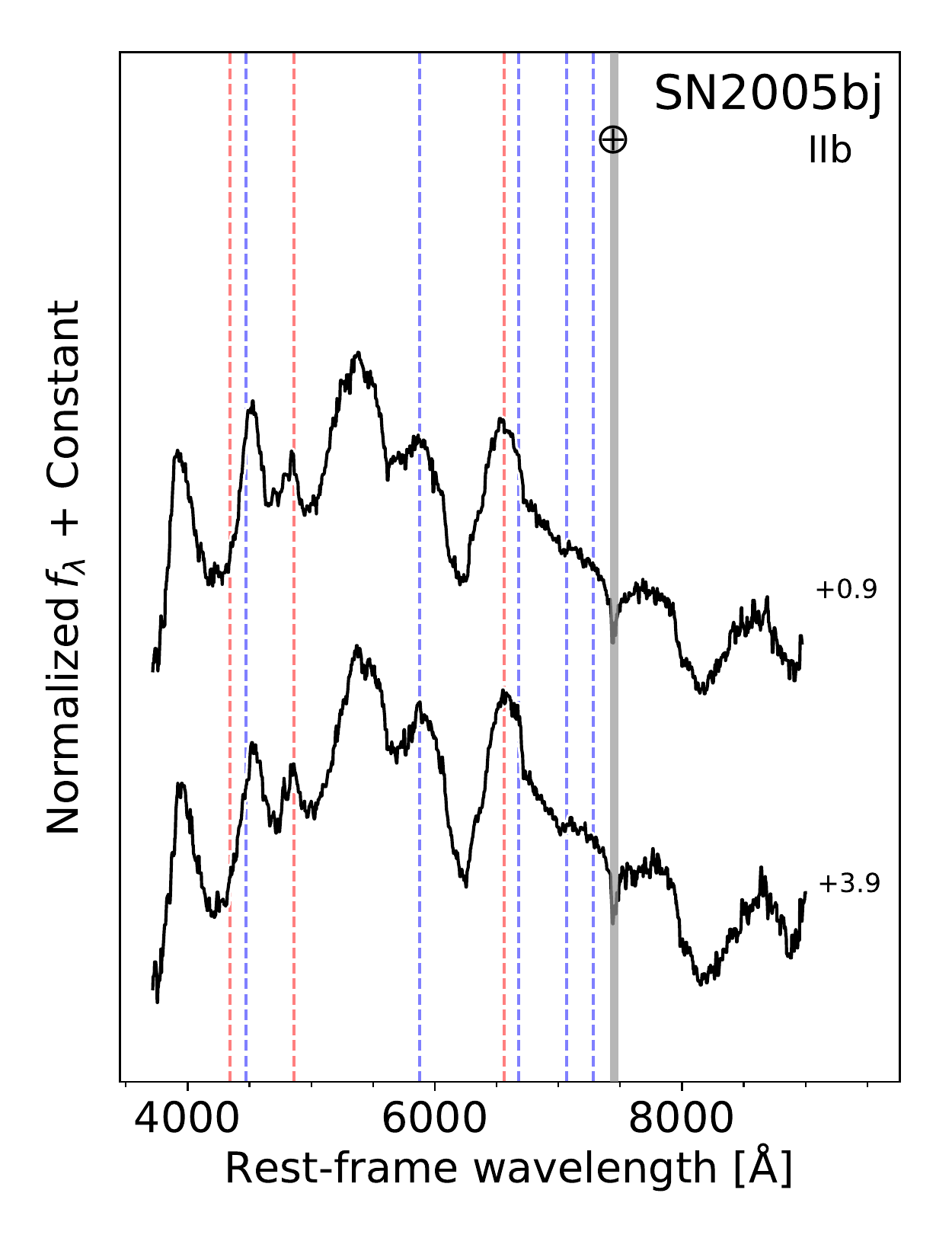} &
    \includegraphics[width=.45\linewidth]{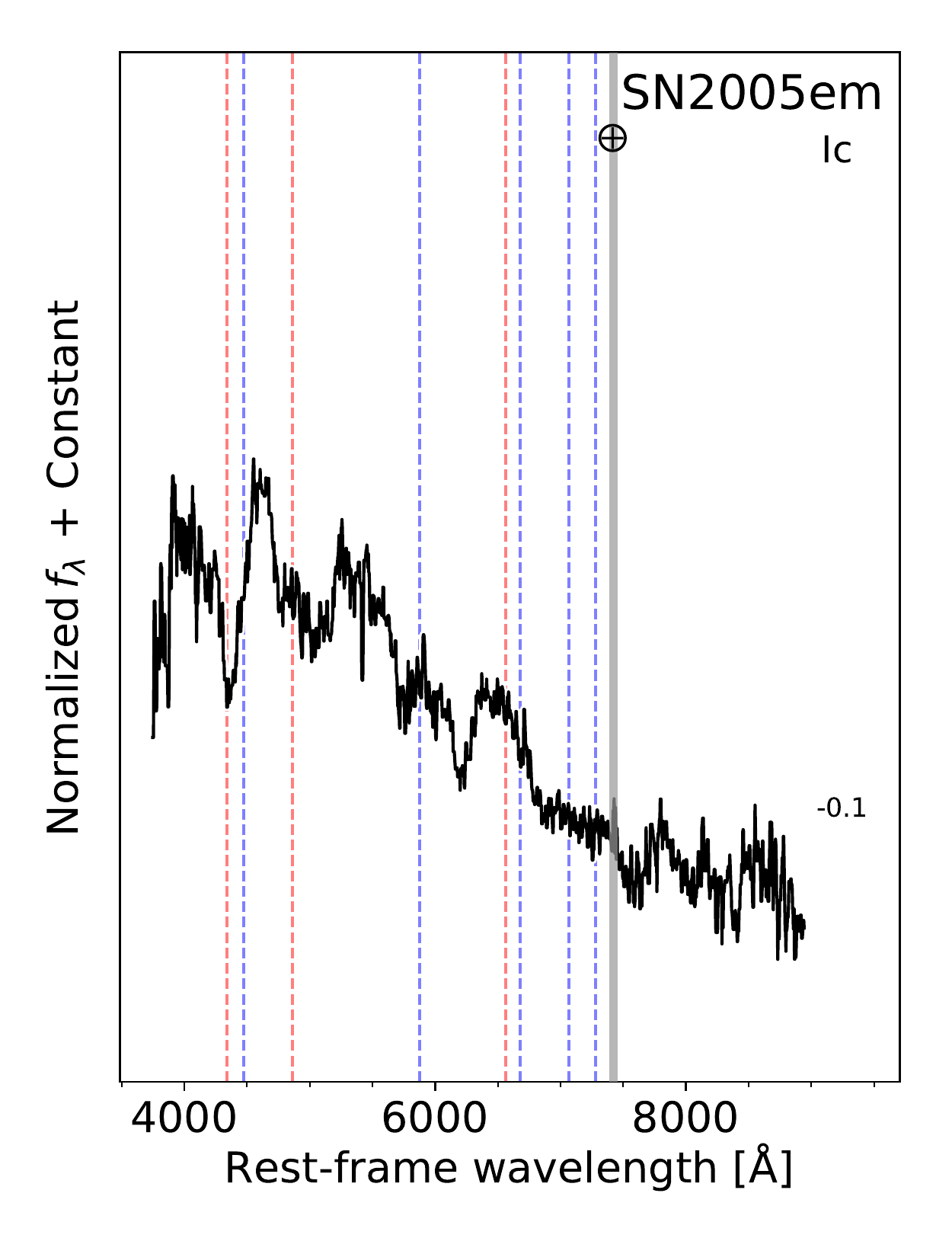} \\
  \end{array}$
  \caption{Continued.}
\end{figure*}

\setcounter{figure}{1}
\begin{figure*}
	\setlength\arraycolsep{0pt}
	\renewcommand{\arraystretch}{0}
	\centering$
  \begin{array}{cc}
    \includegraphics[width=.45\linewidth]{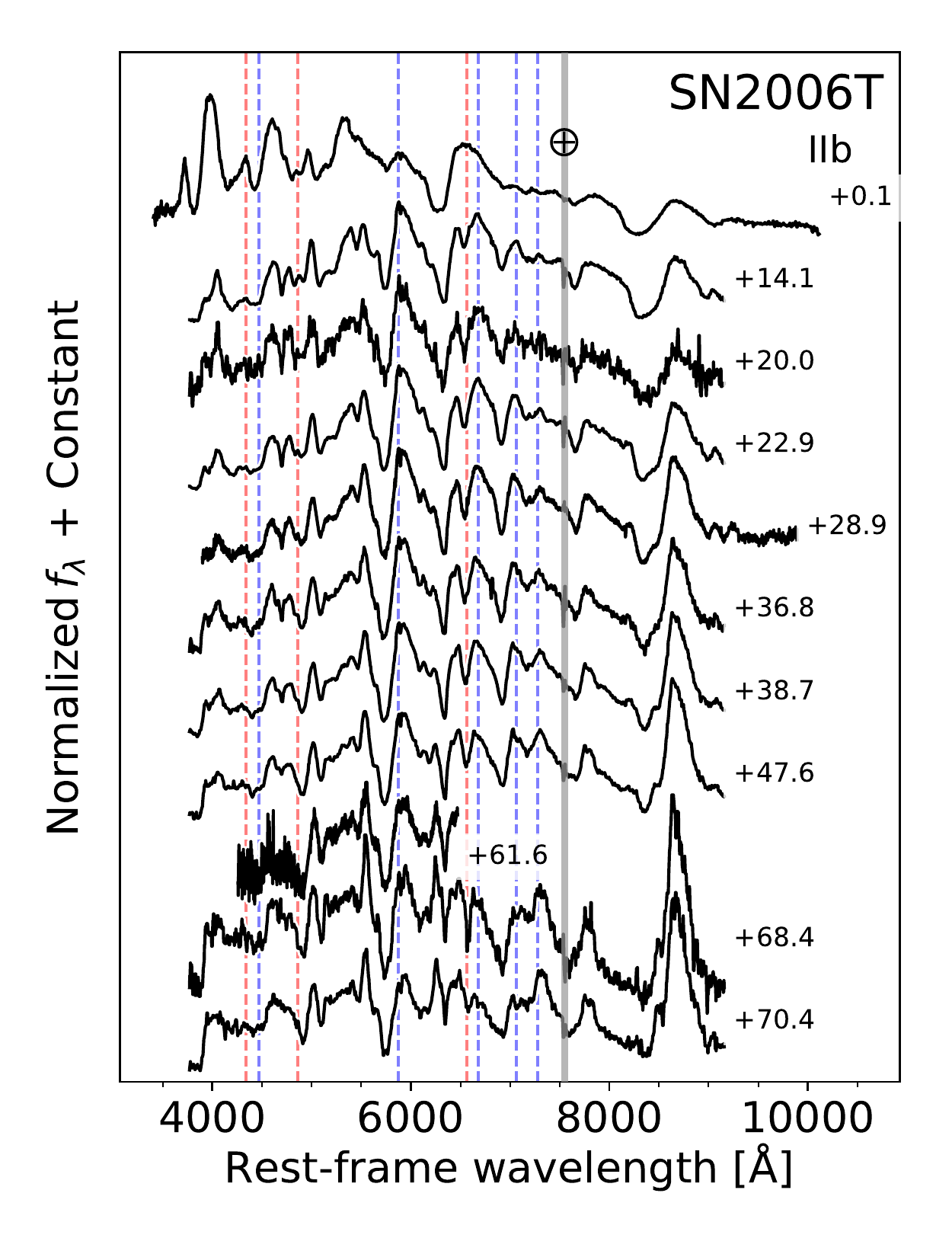} &
    \includegraphics[width=.45\linewidth]{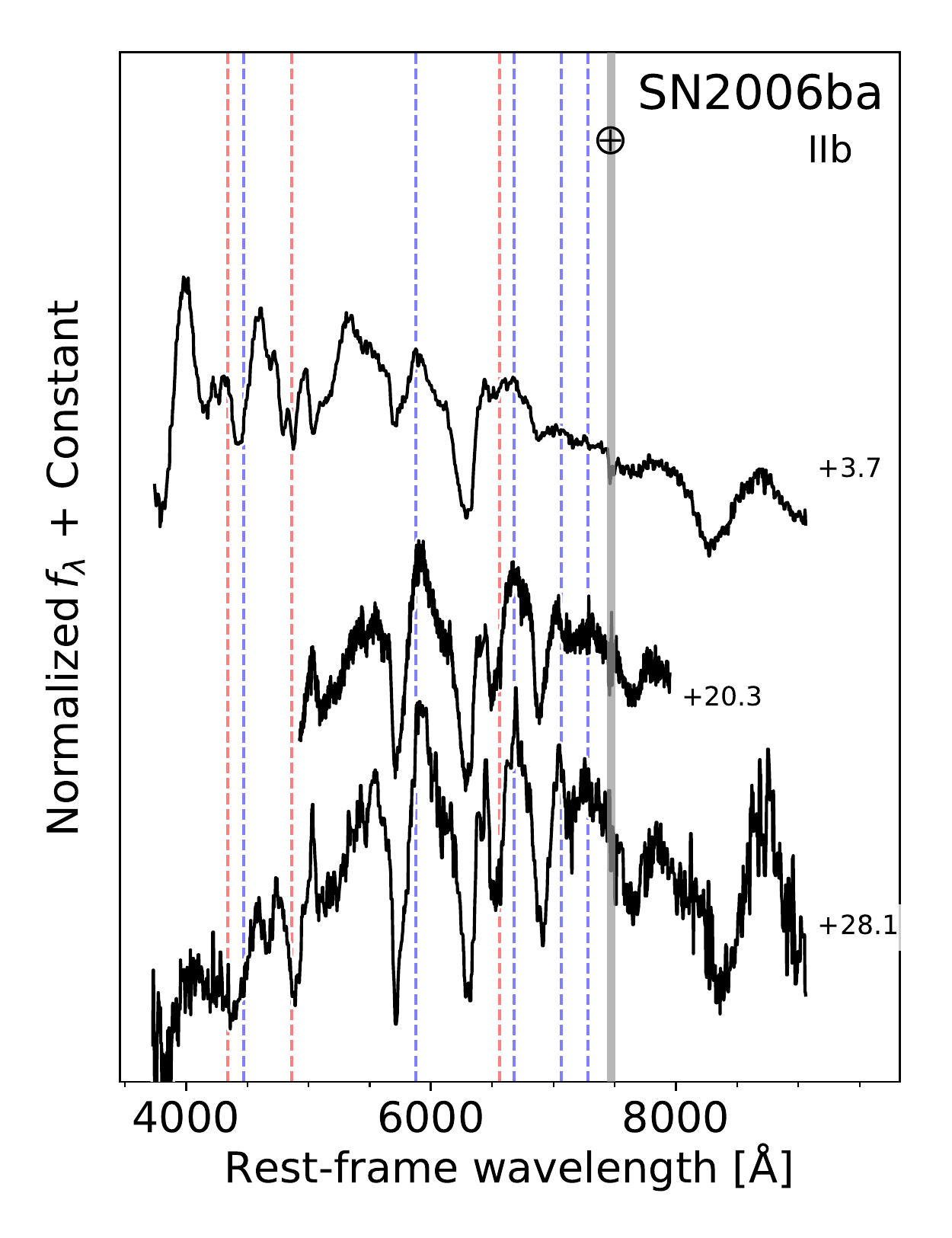} \\
    \includegraphics[width=.45\linewidth]{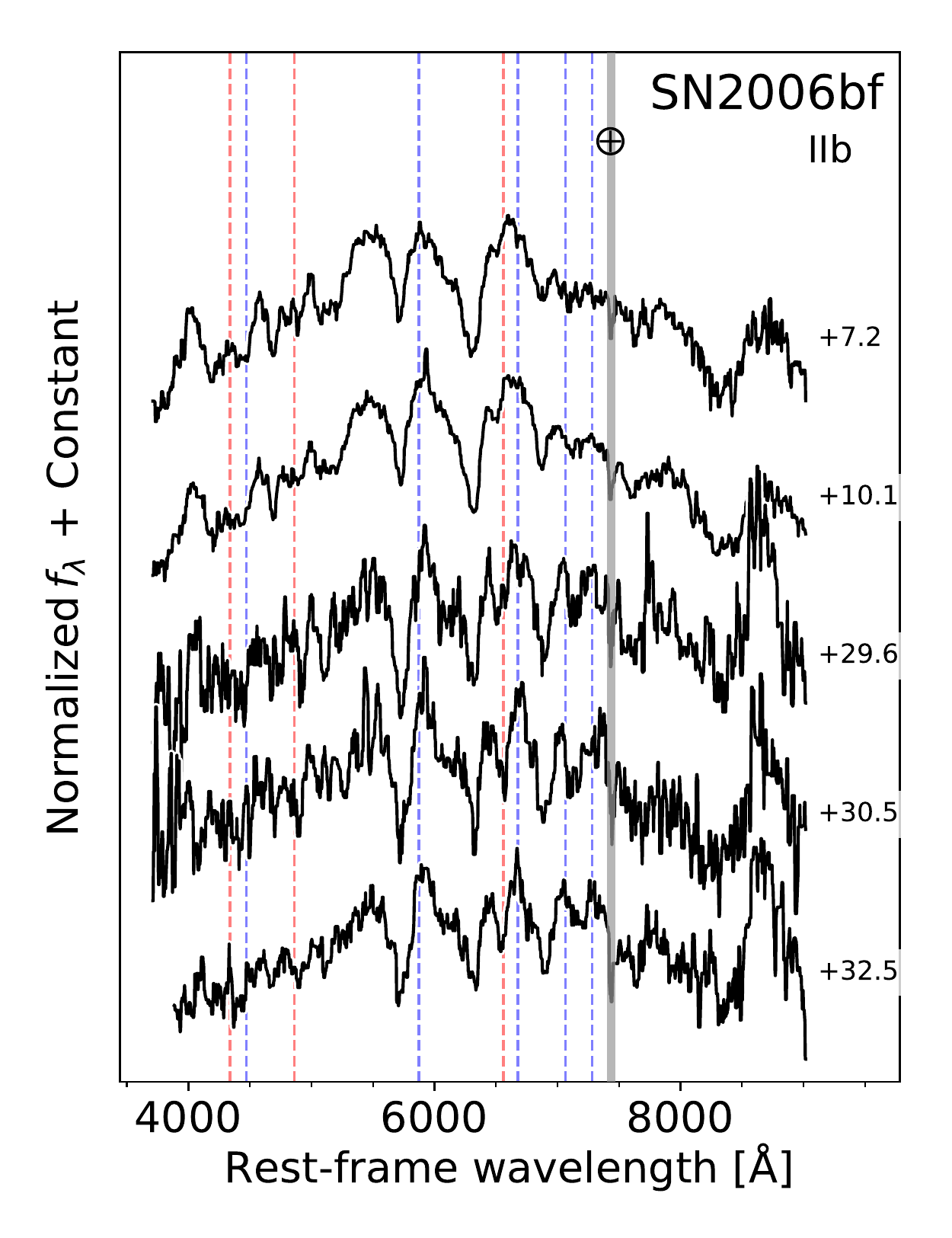} &
    \includegraphics[width=.45\linewidth]{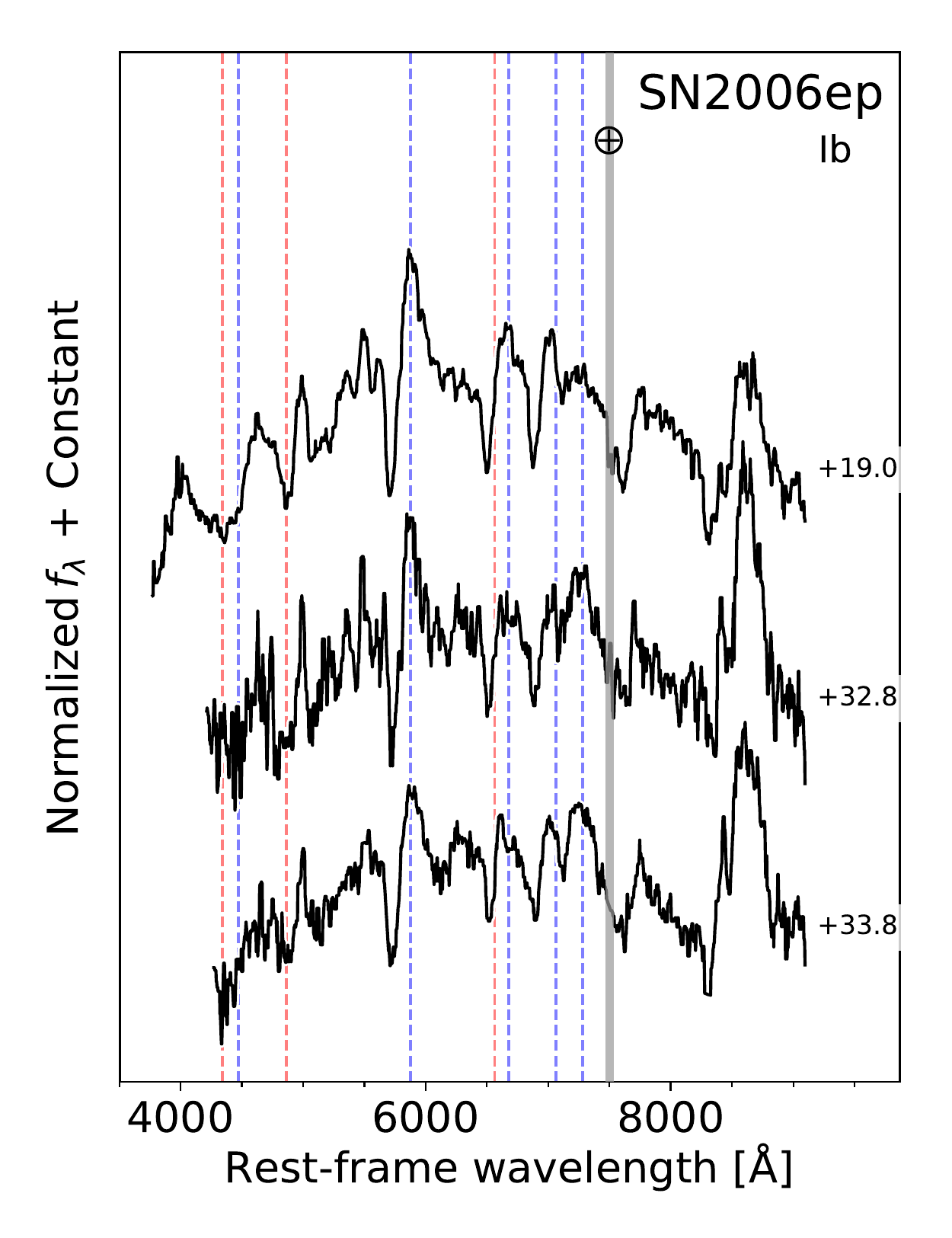} \\
  \end{array}$
   \caption{Continued.}
\end{figure*}

\setcounter{figure}{1}
\begin{figure*}
	\setlength\arraycolsep{0pt}
	\renewcommand{\arraystretch}{0}
	\centering$
  \begin{array}{cc}
    \includegraphics[width=.45\linewidth]{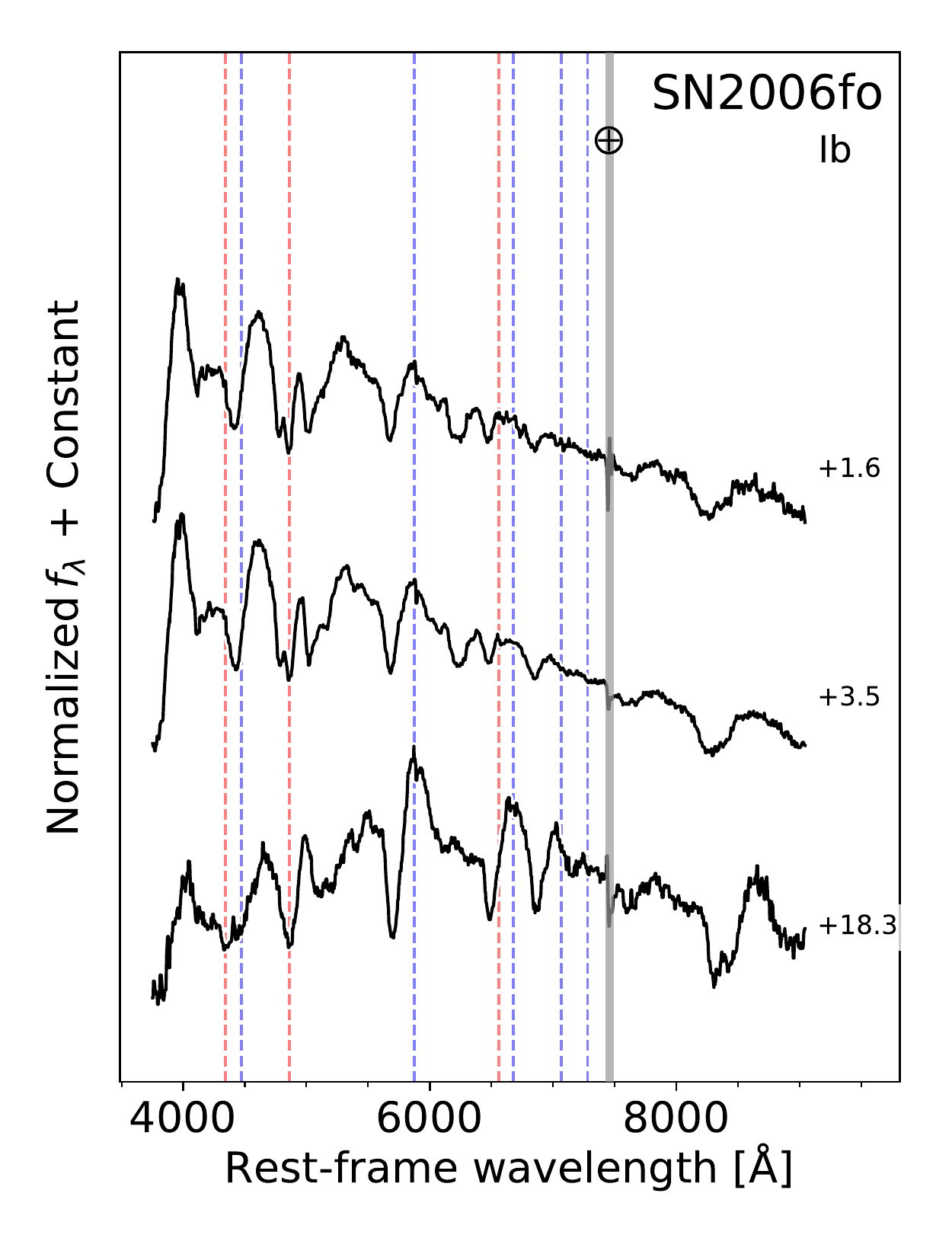} &
    \includegraphics[width=.45\linewidth]{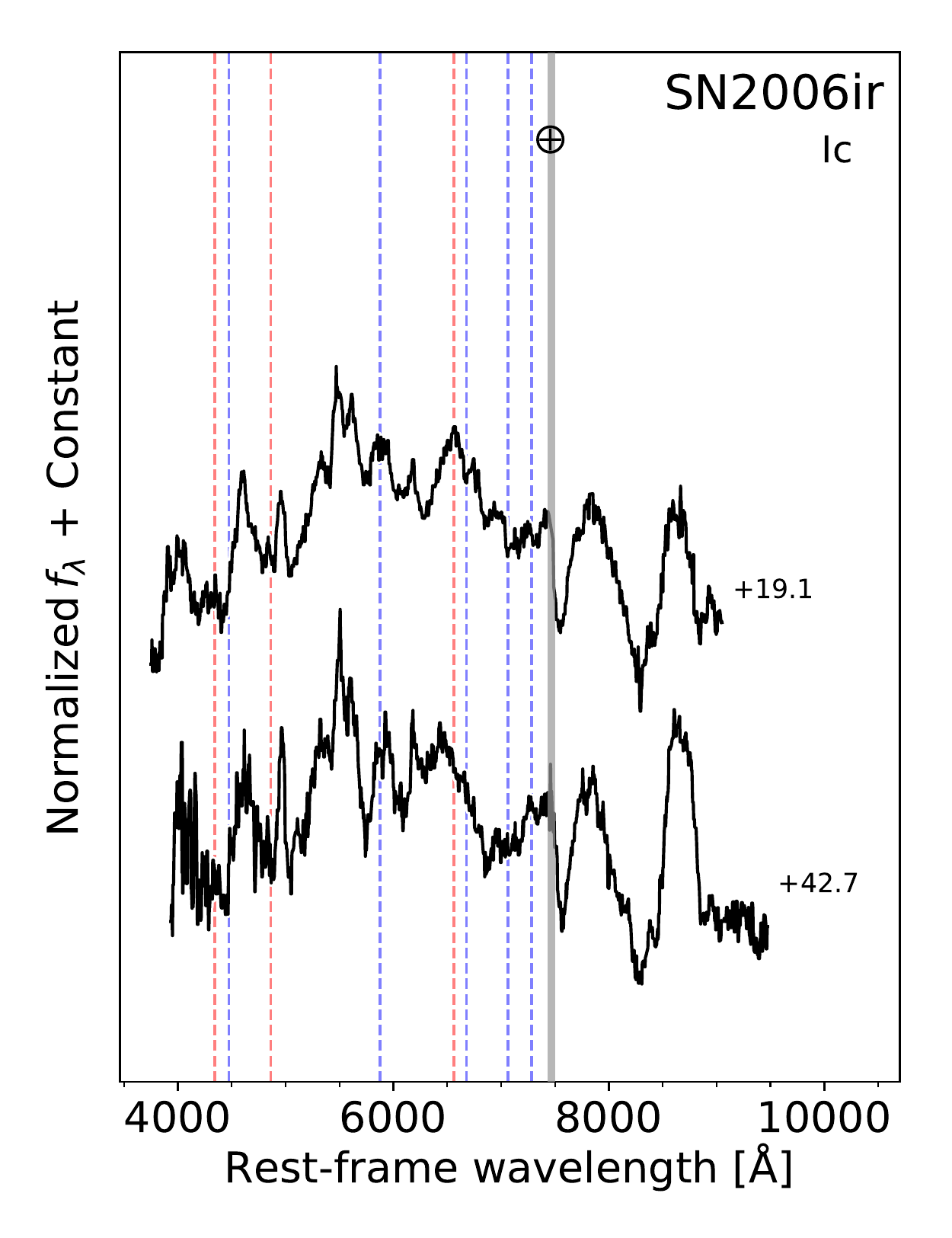} \\
    \includegraphics[width=.45\linewidth]{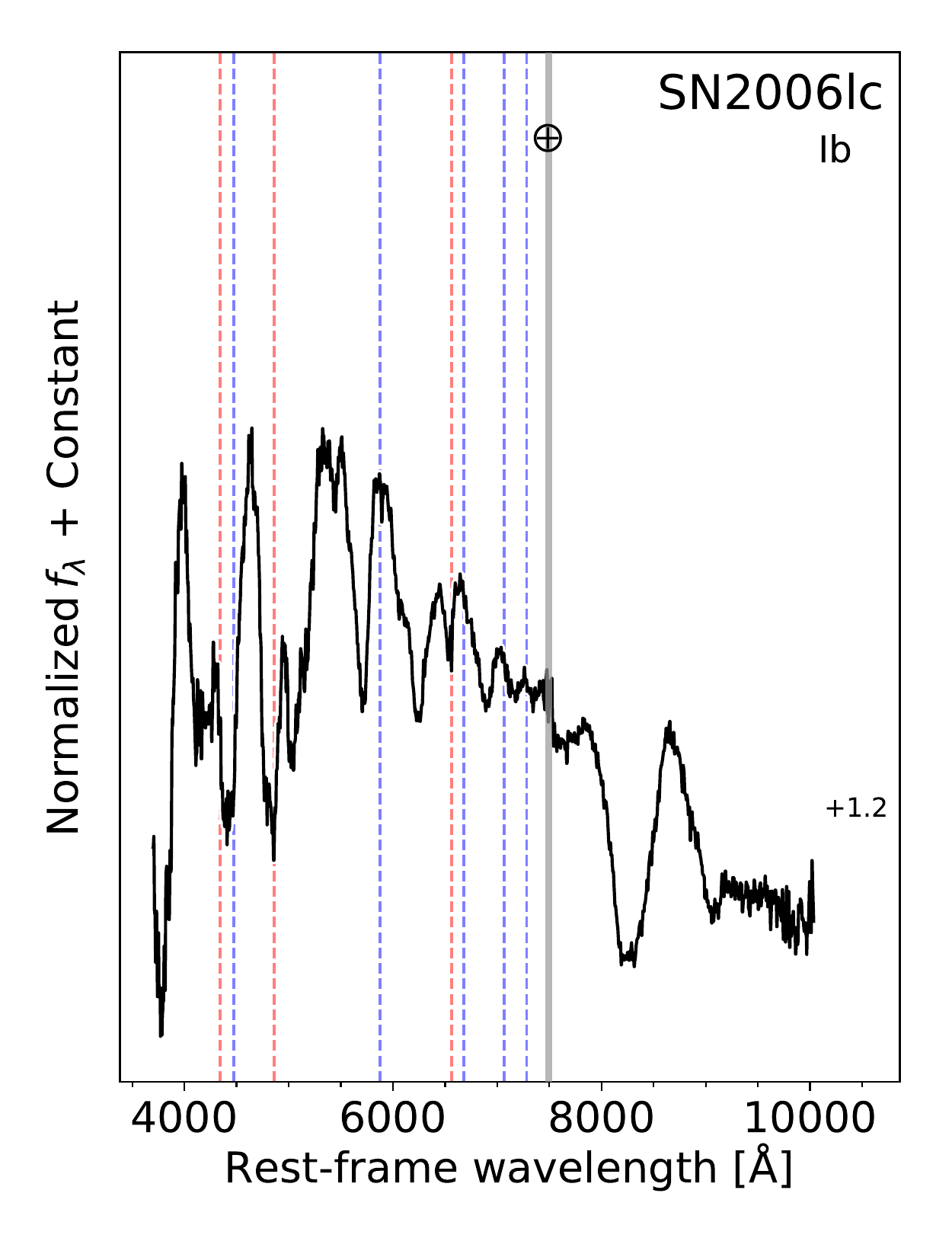} &
    \includegraphics[width=.45\linewidth]{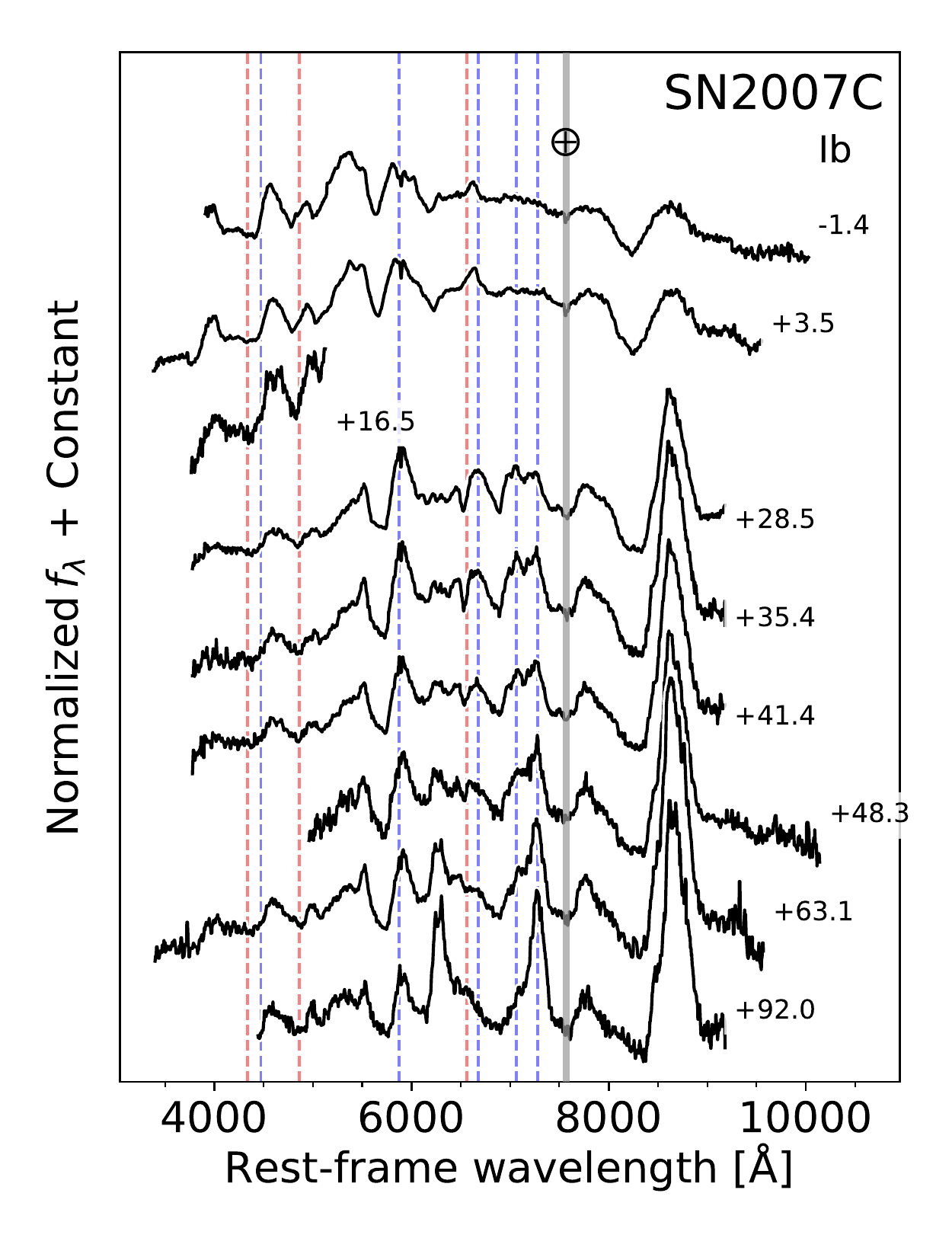} \\
  \end{array}$
   \caption{Continued.}
\end{figure*}

\setcounter{figure}{1}
\begin{figure*}
	\setlength\arraycolsep{0pt}
	\renewcommand{\arraystretch}{0}
	\centering$
  \begin{array}{cc}
    \includegraphics[width=.45\linewidth]{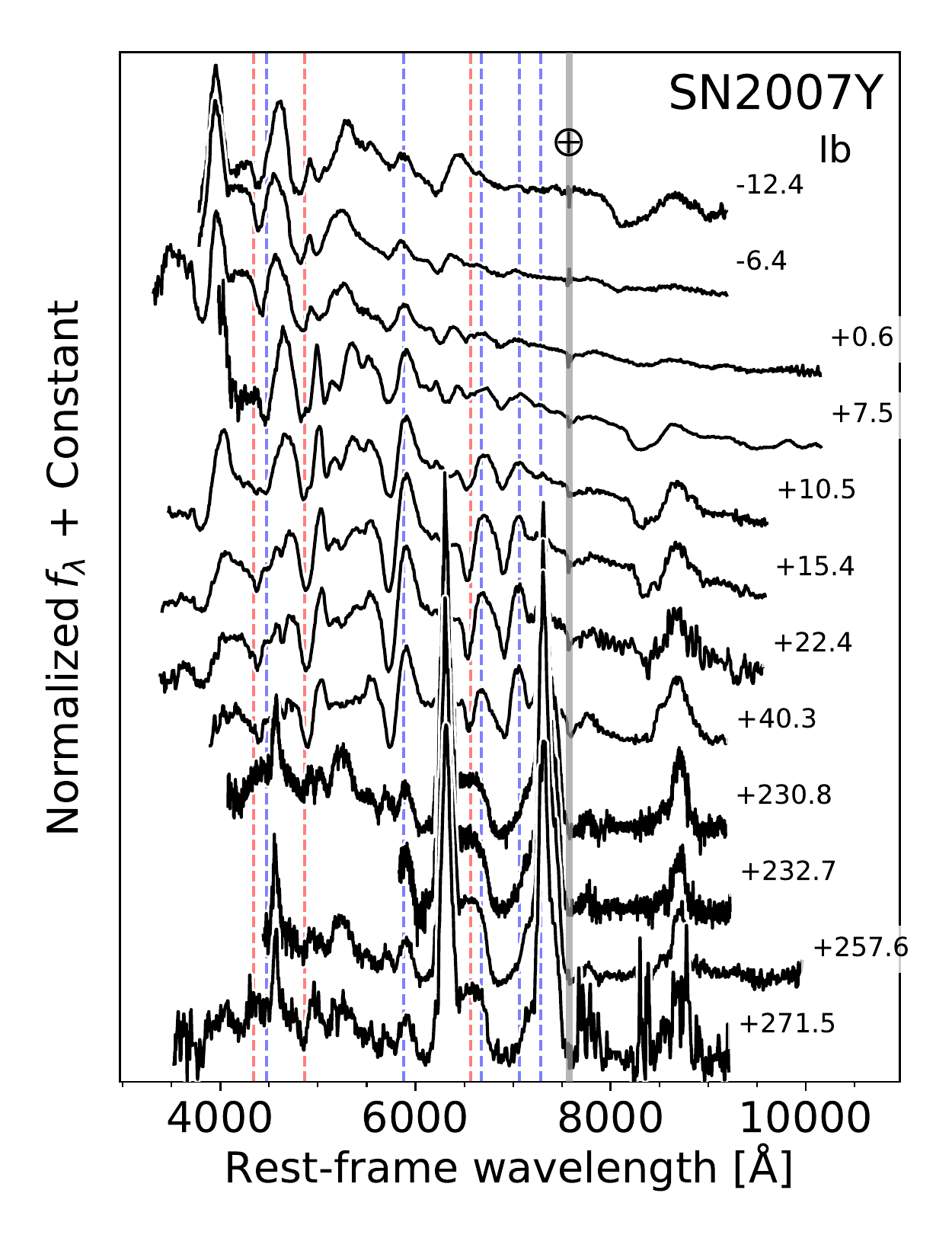} & 
    \includegraphics[width=.45\linewidth]{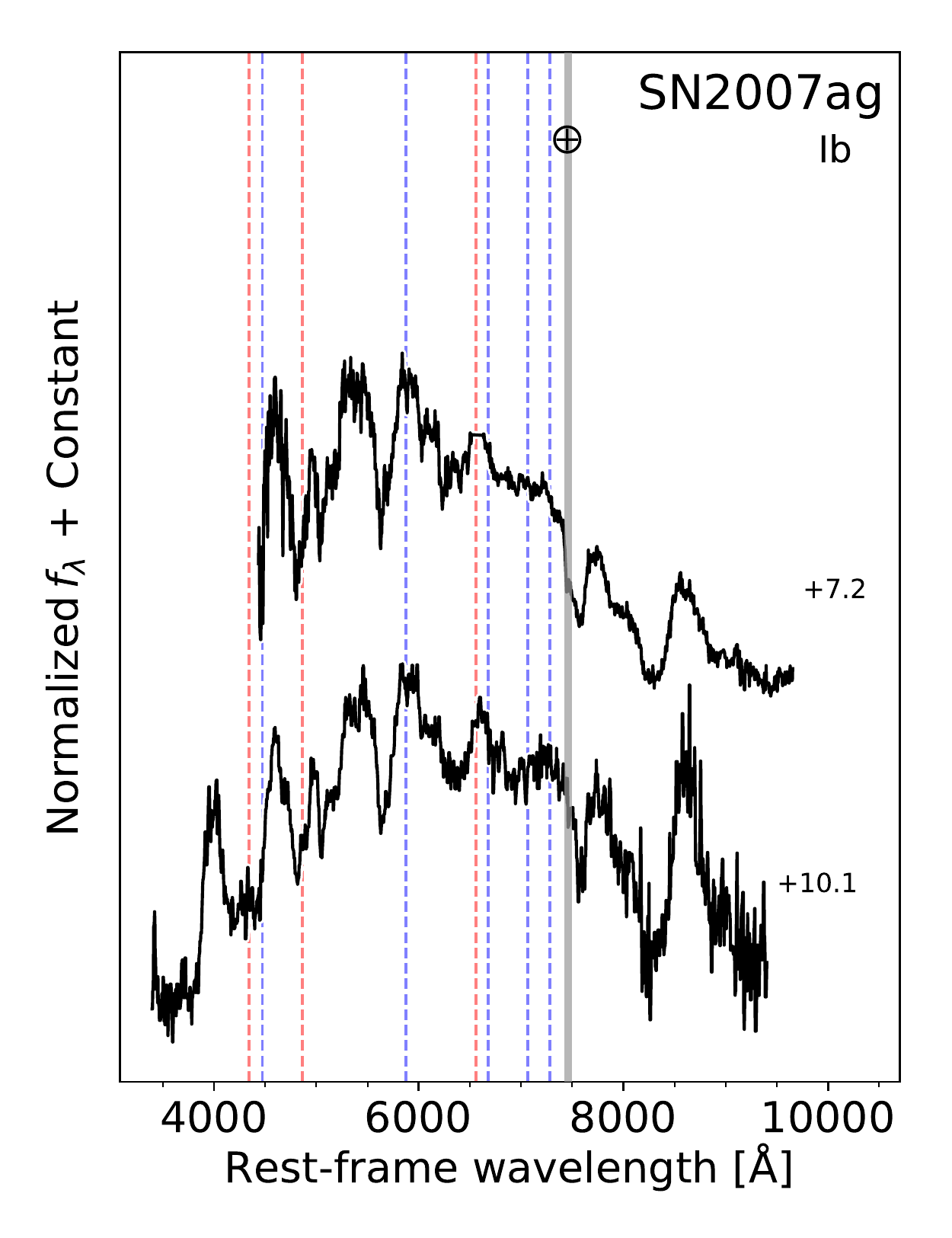} \\
    \includegraphics[width=.45\linewidth]{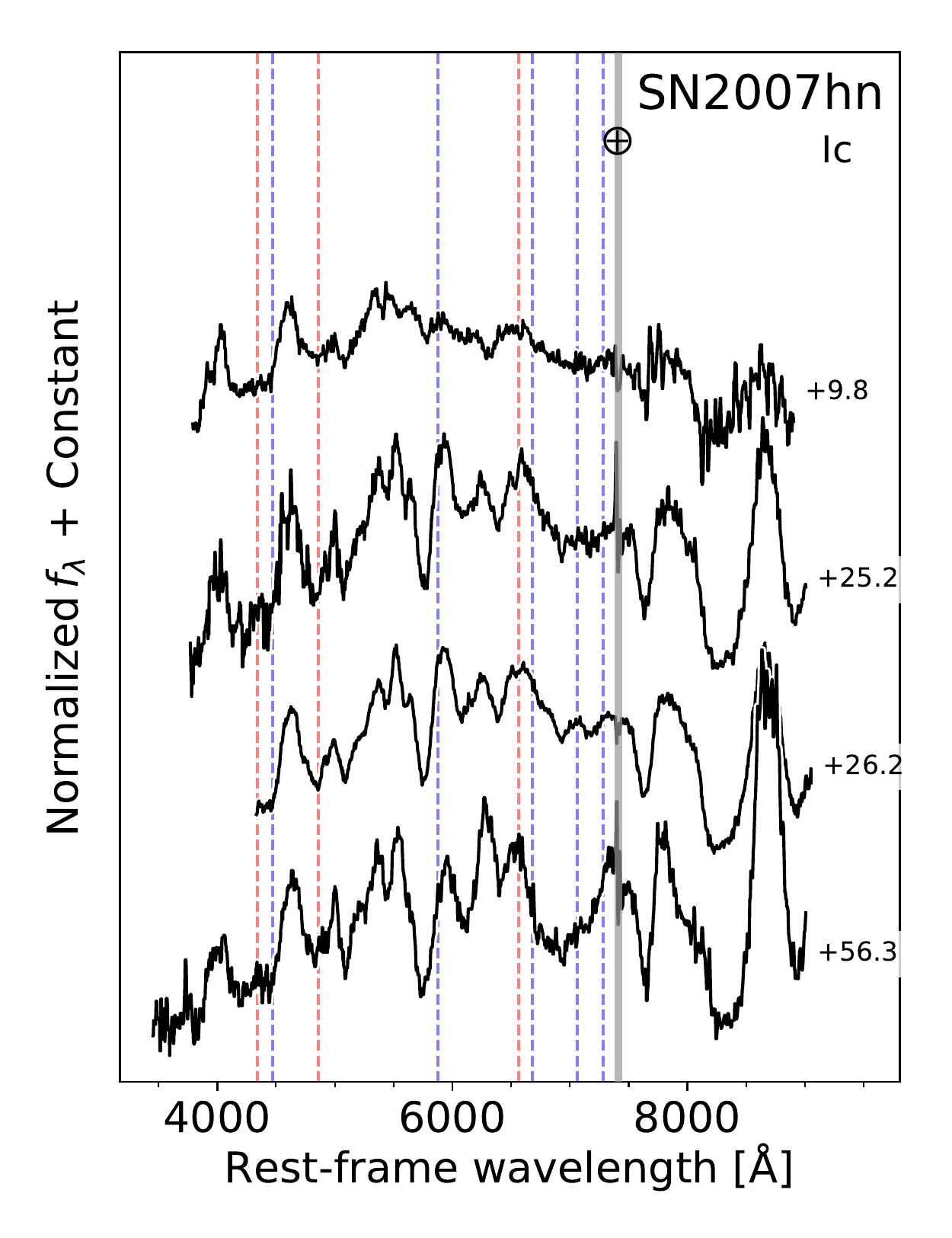} &
    \includegraphics[width=.45\linewidth]{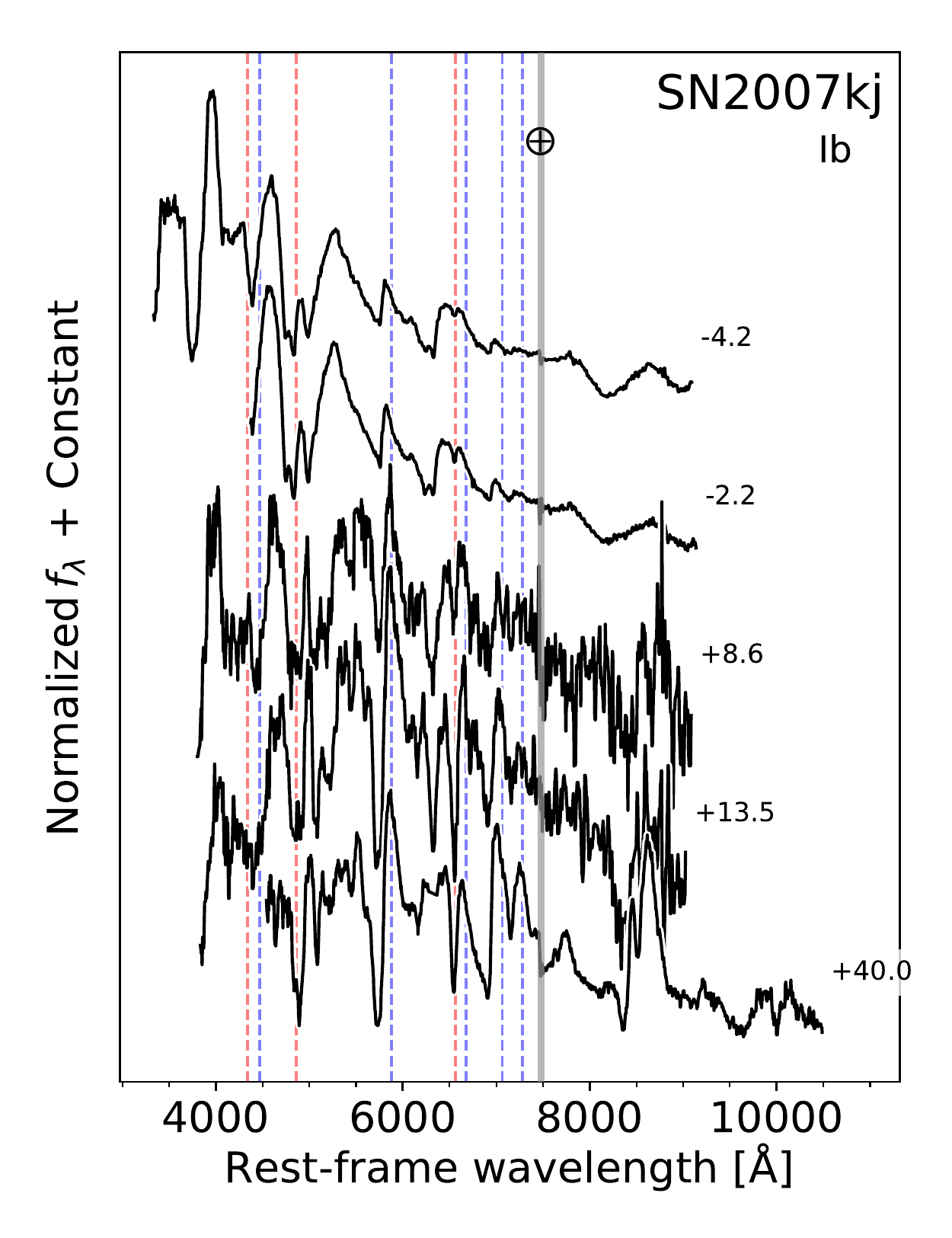} \\
  \end{array}$
   \caption{Continued.}
\end{figure*}

\setcounter{figure}{1}
\begin{figure*}
	\setlength\arraycolsep{0pt}
	\renewcommand{\arraystretch}{0}
	\centering$
  \begin{array}{cc}
    \includegraphics[width=.45\linewidth]{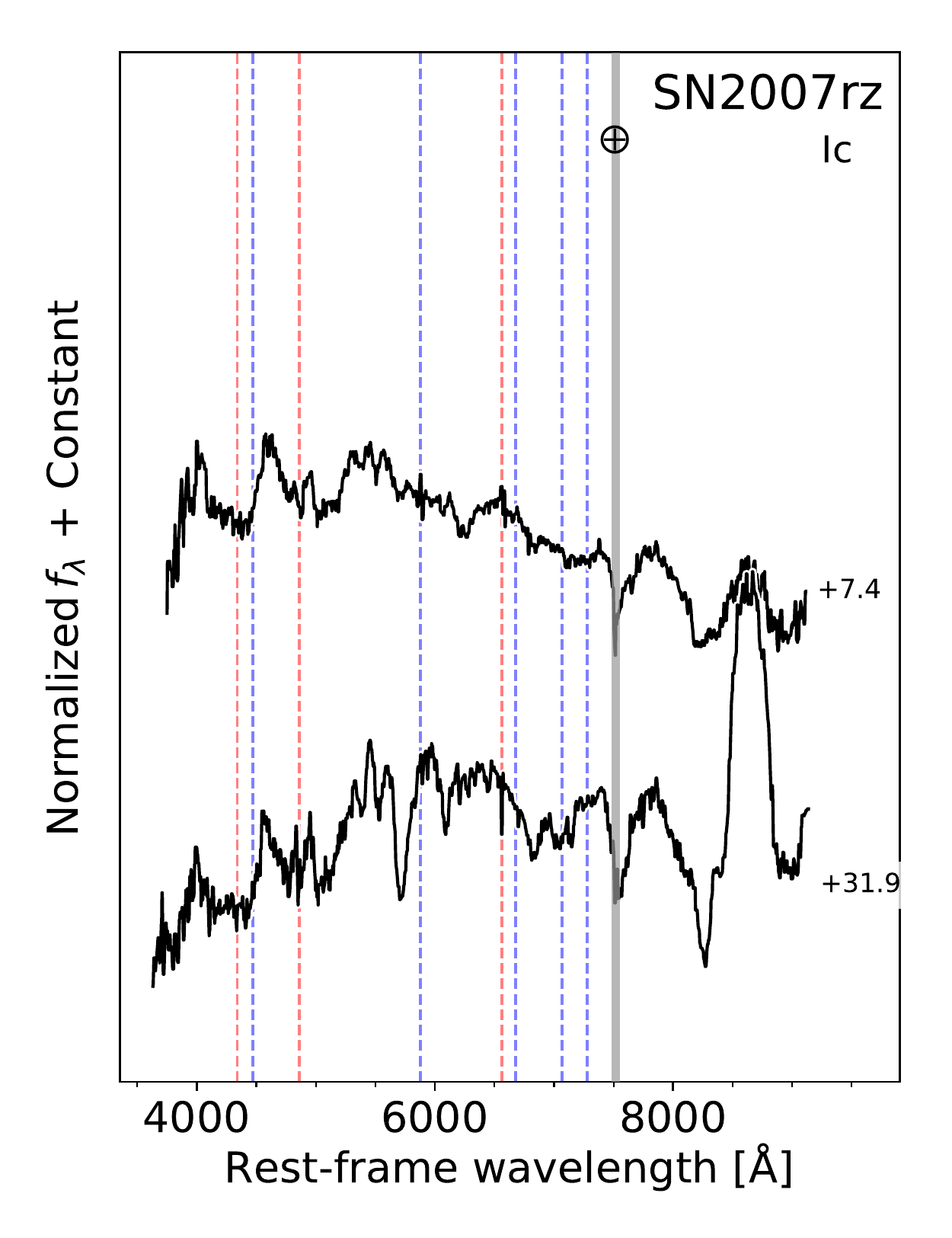} &
    \includegraphics[width=.45\linewidth]{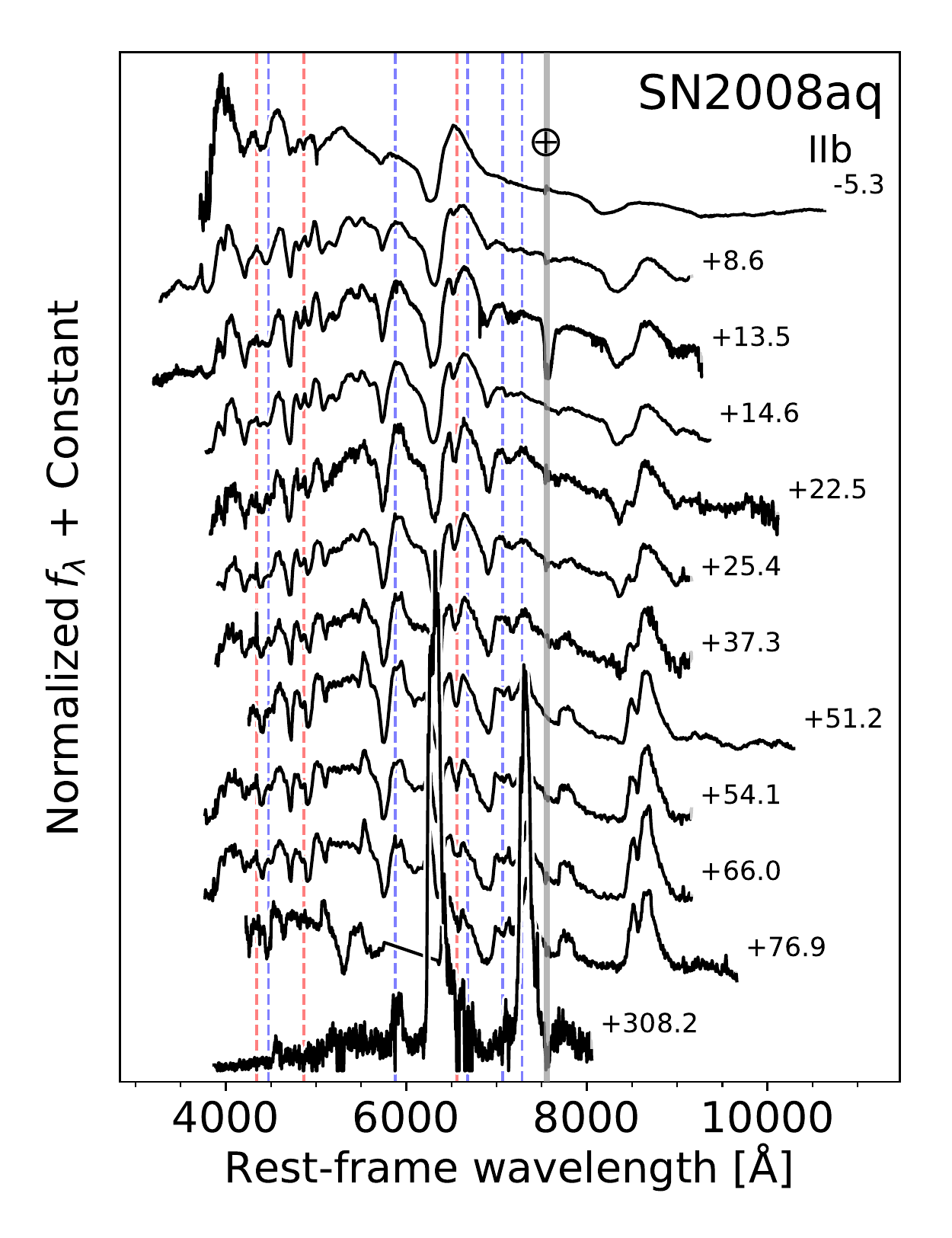} \\
    \includegraphics[width=.45\linewidth]{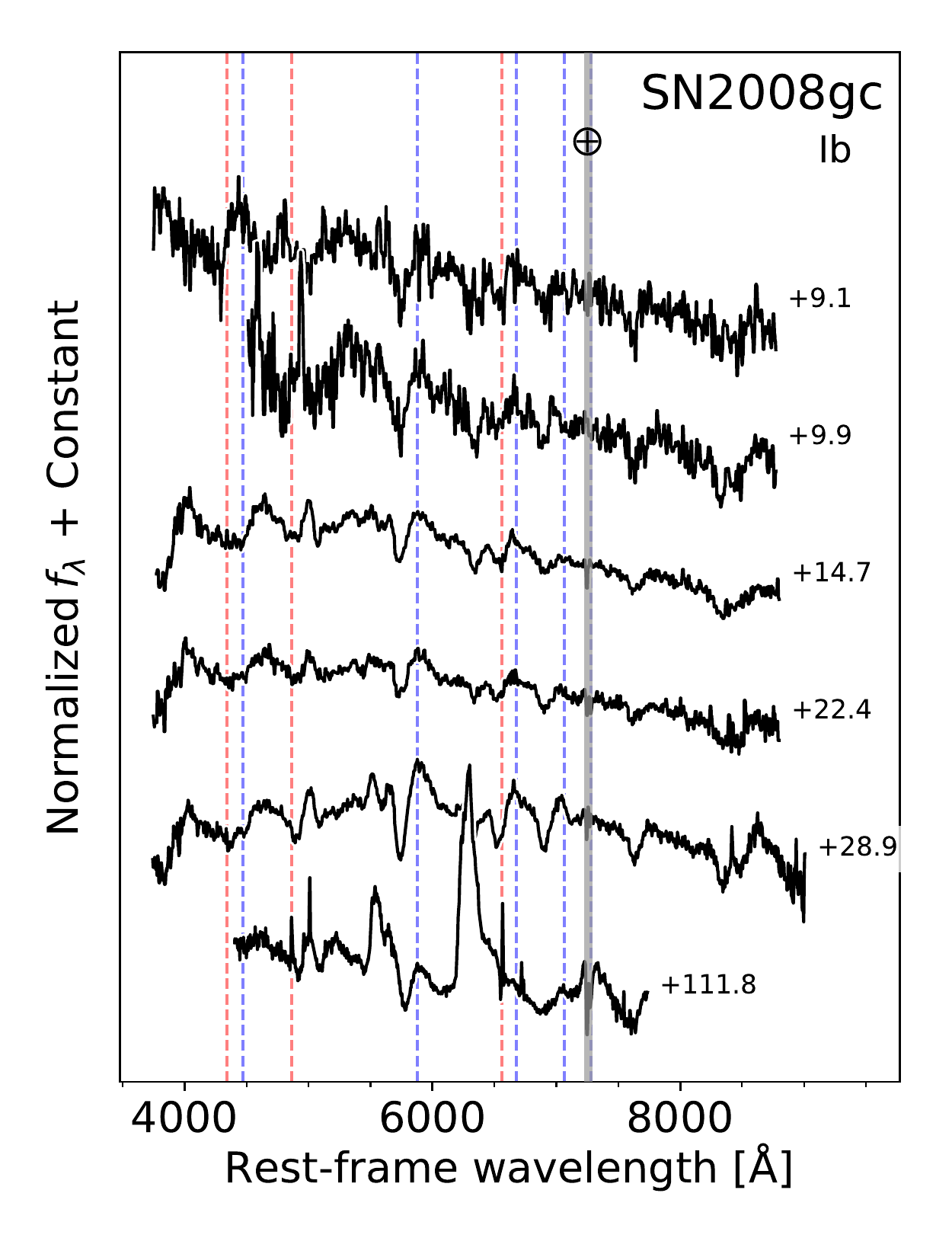} &
    \includegraphics[width=.45\linewidth]{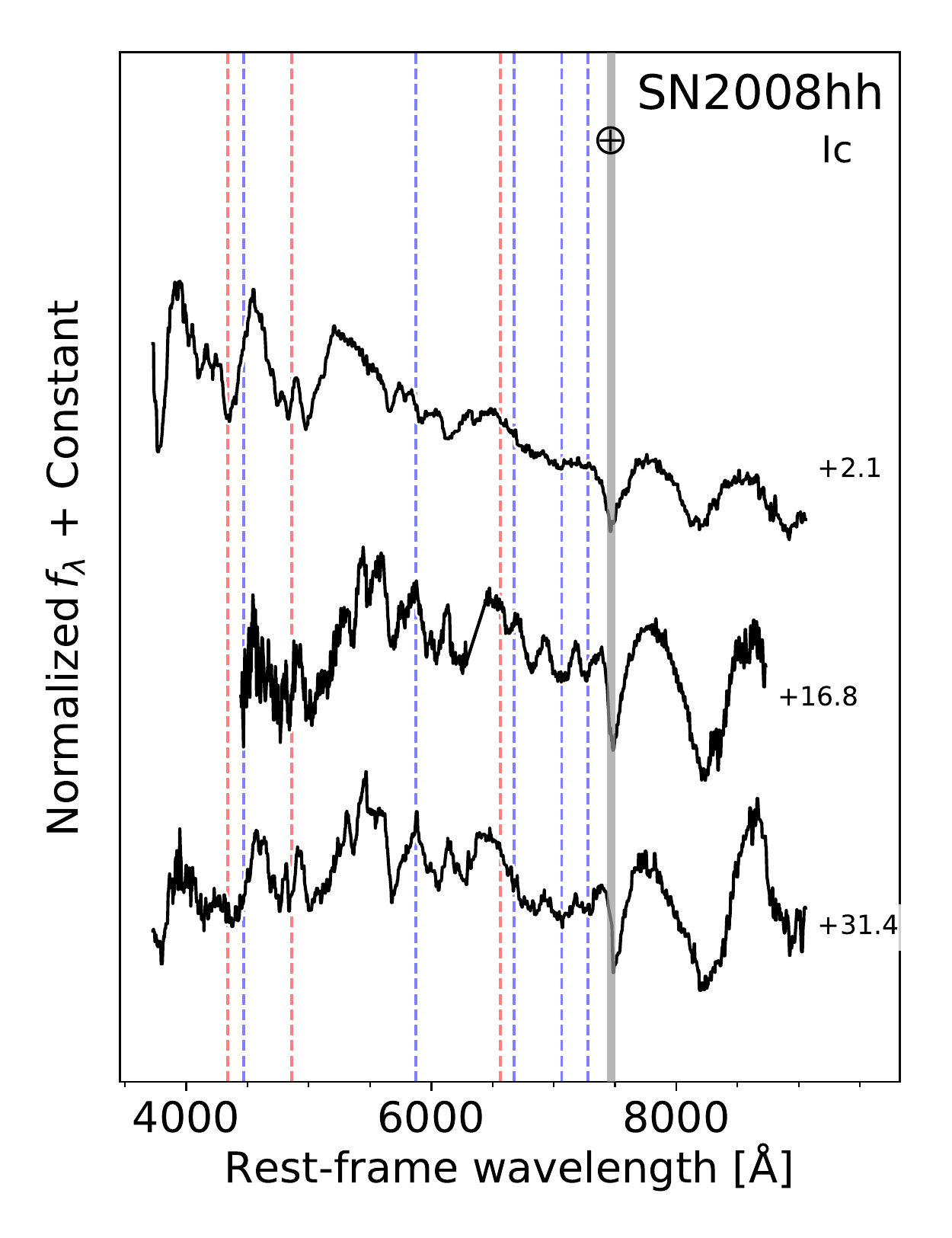} \\
  \end{array}$
 \caption{Continued.}
 \end{figure*}
 
\setcounter{figure}{1}
\begin{figure*}
	\setlength\arraycolsep{0pt}
	\renewcommand{\arraystretch}{0}
	\centering$
  \begin{array}{cc}
       \includegraphics[width=.45\linewidth]{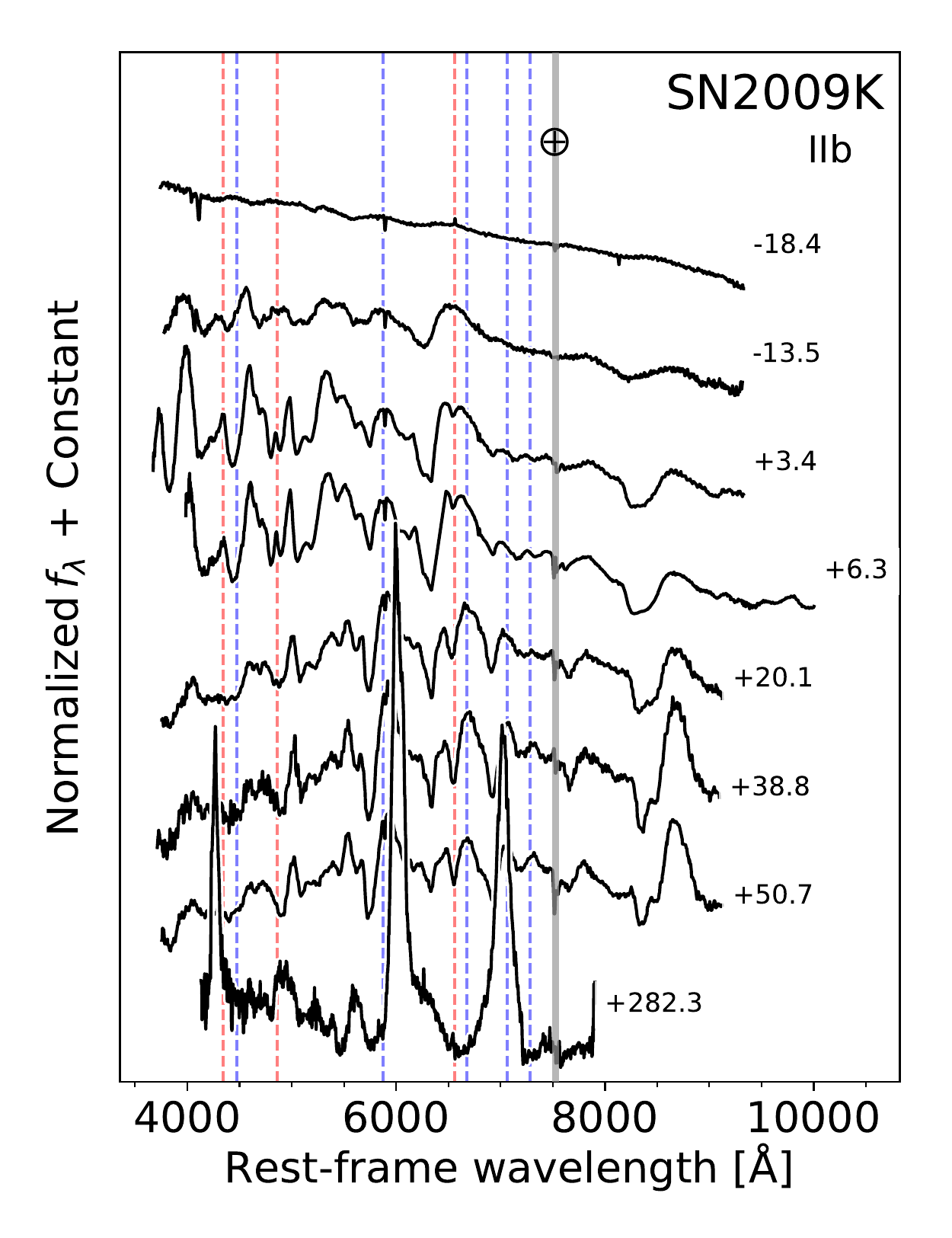} &
    \includegraphics[width=.45\linewidth]{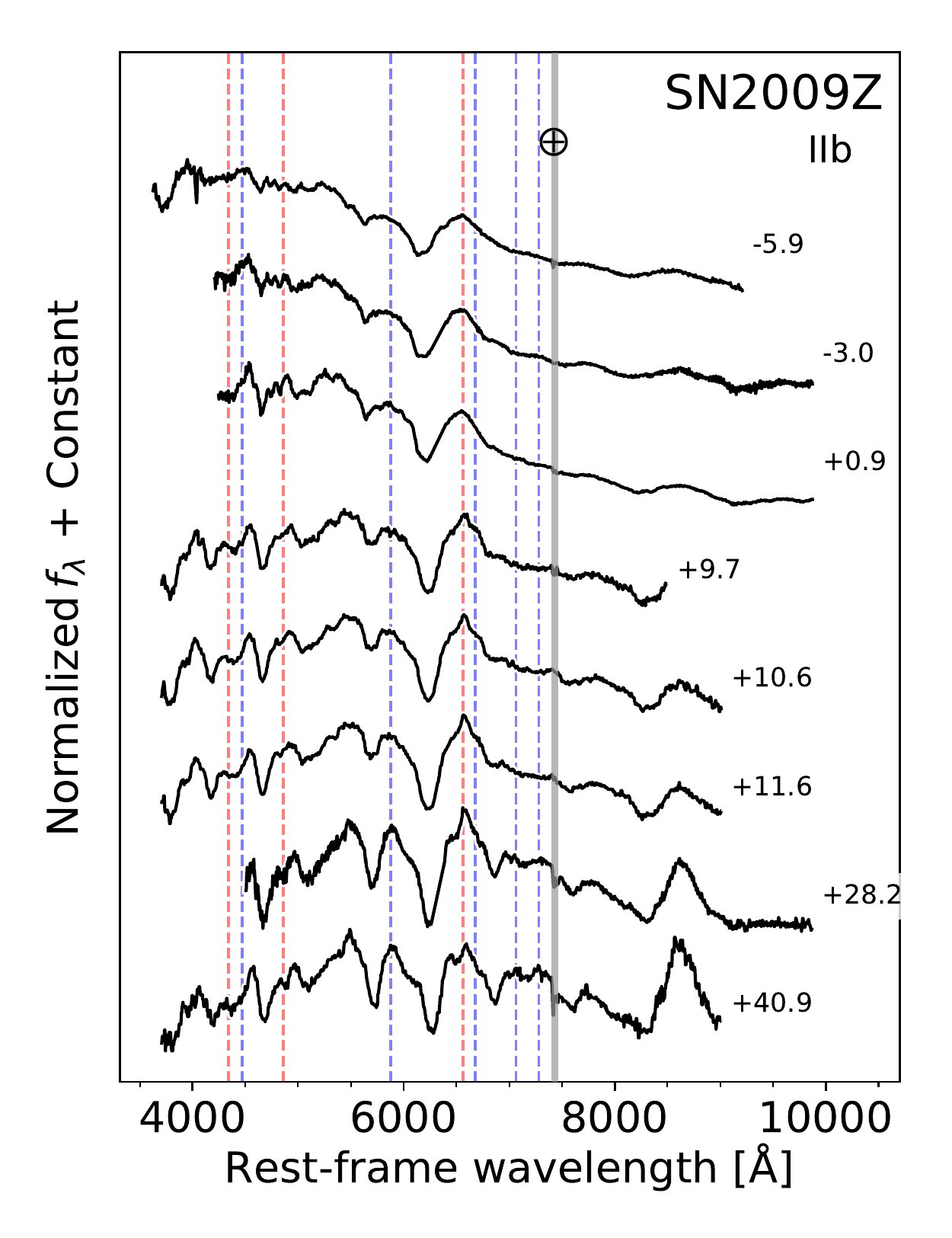} \\
    \includegraphics[width=.45\linewidth]{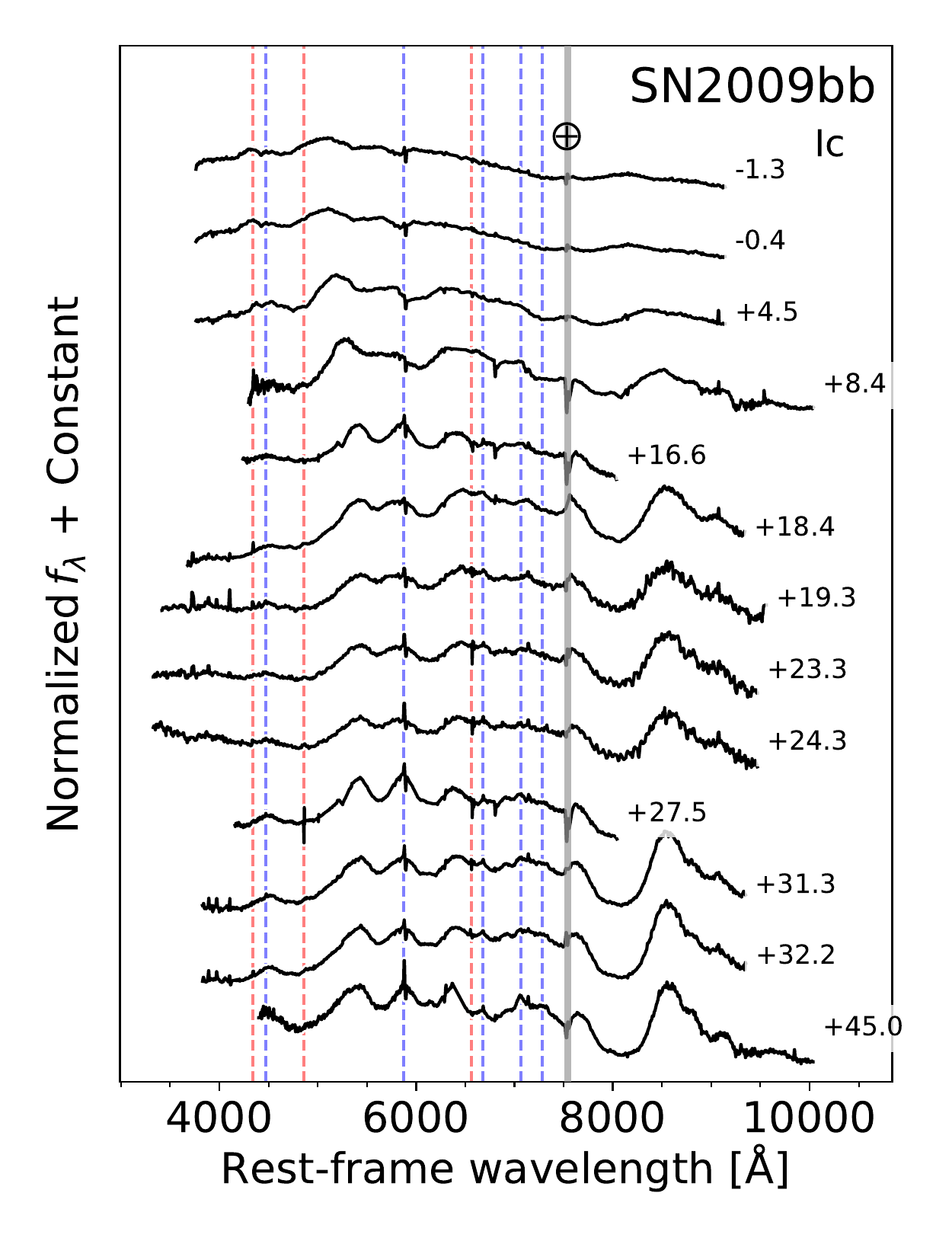} &
    \includegraphics[width=.45\linewidth]{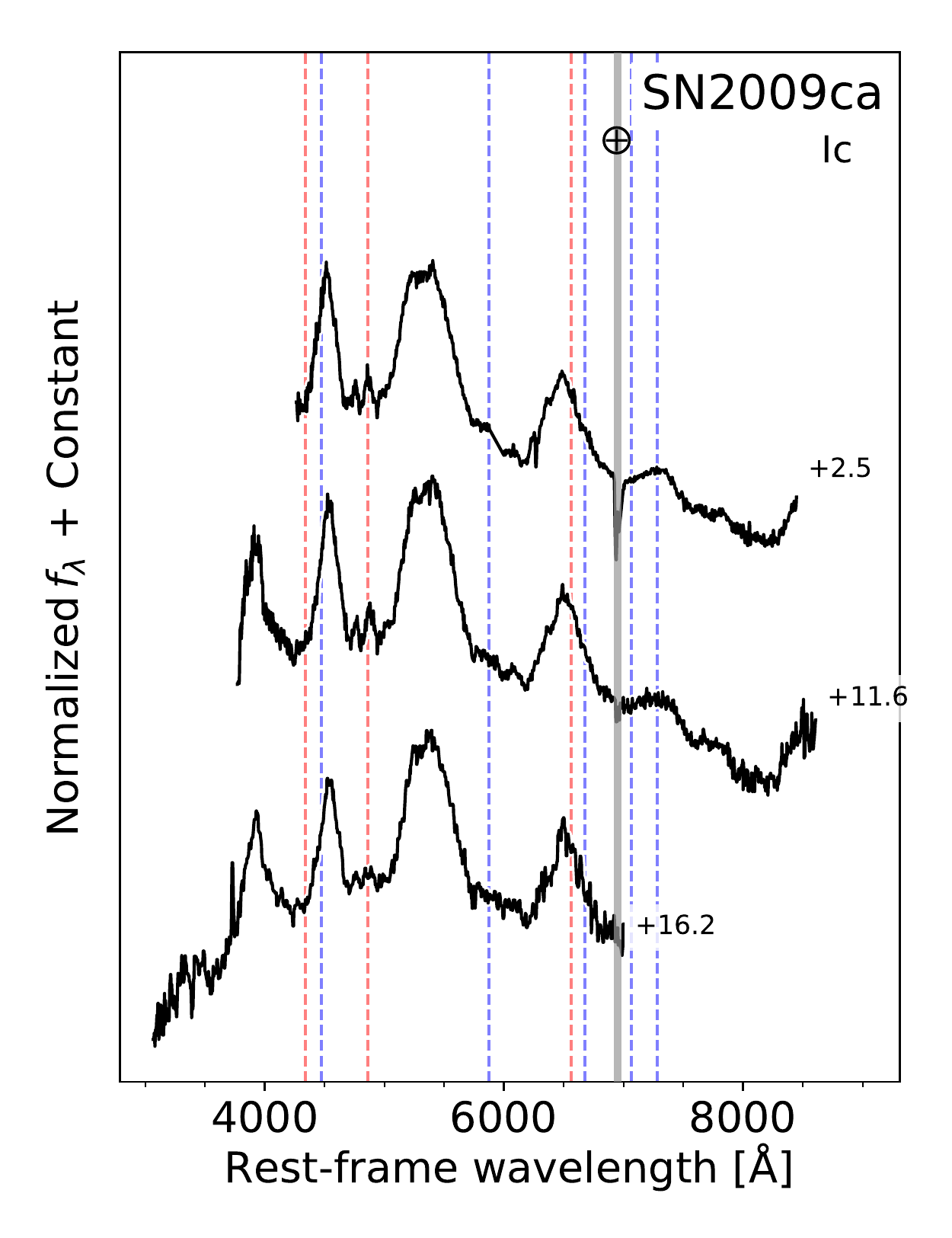} \\
  \end{array}$
   \caption{Continued.}
\end{figure*}

\setcounter{figure}{1}
\begin{figure*}
	\setlength\arraycolsep{0pt}
	\renewcommand{\arraystretch}{0}
	\centering$
  \begin{array}{cc}
    \includegraphics[width=.45\linewidth]{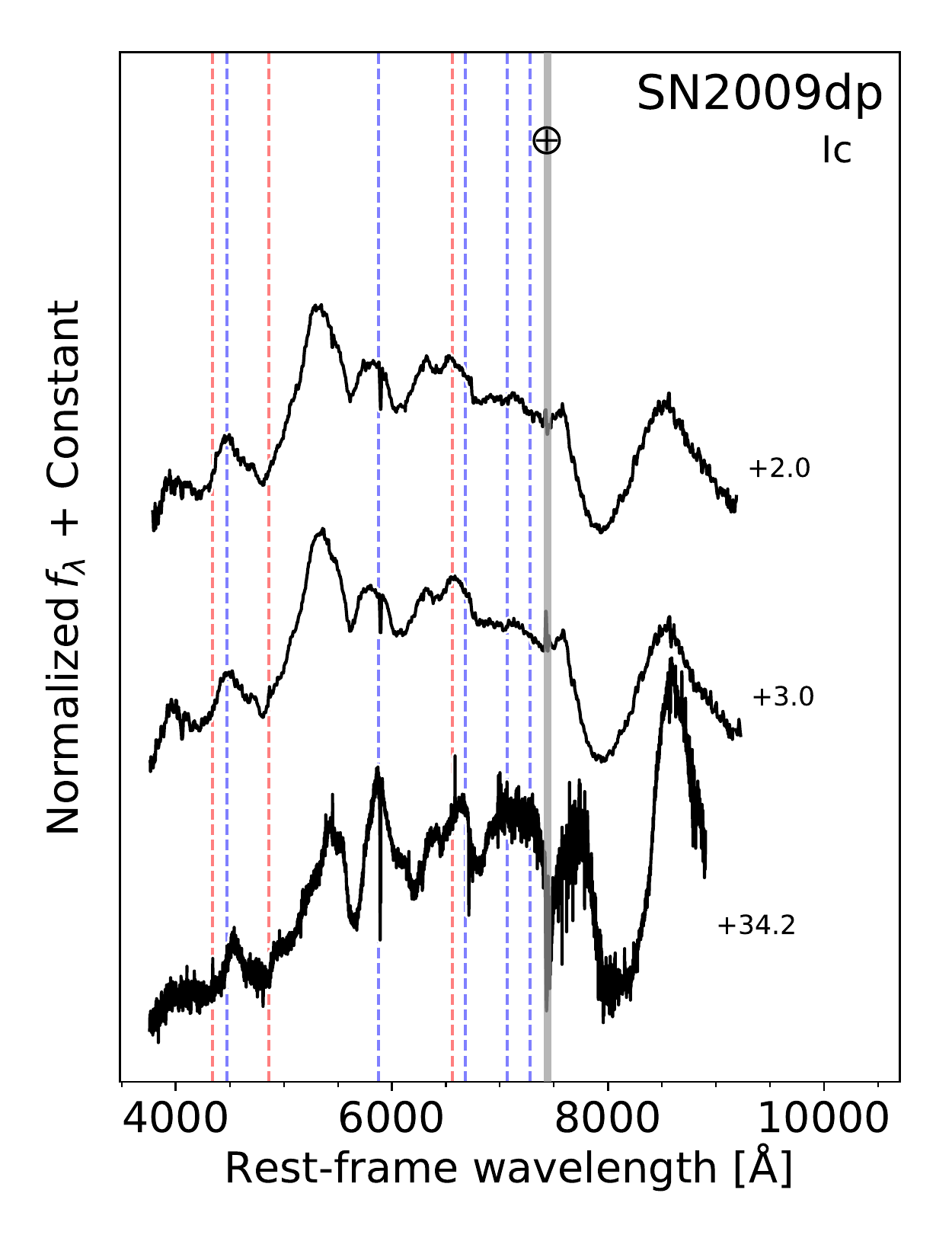} &
    \includegraphics[width=.45\linewidth]{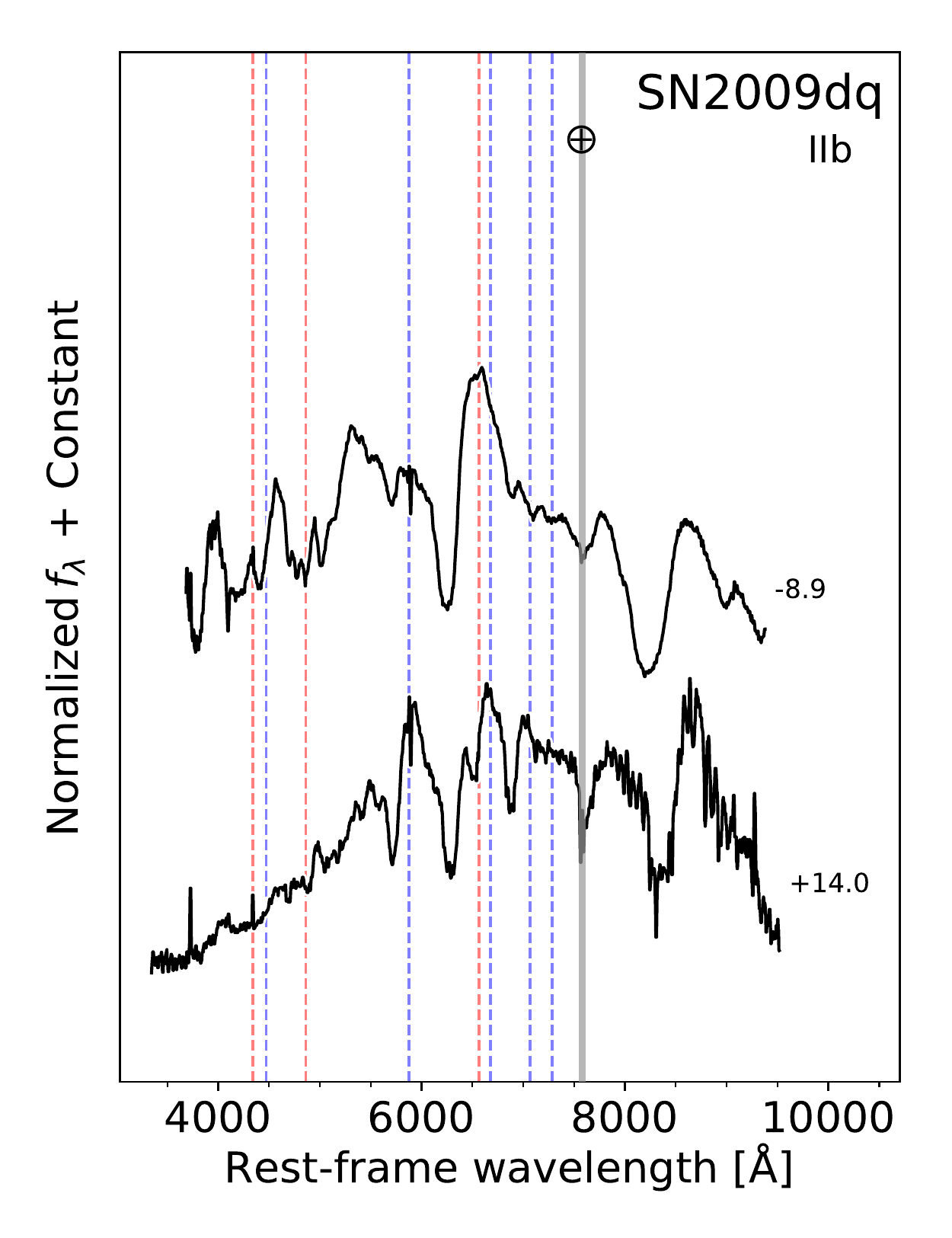} \\
    \includegraphics[width=.45\linewidth]{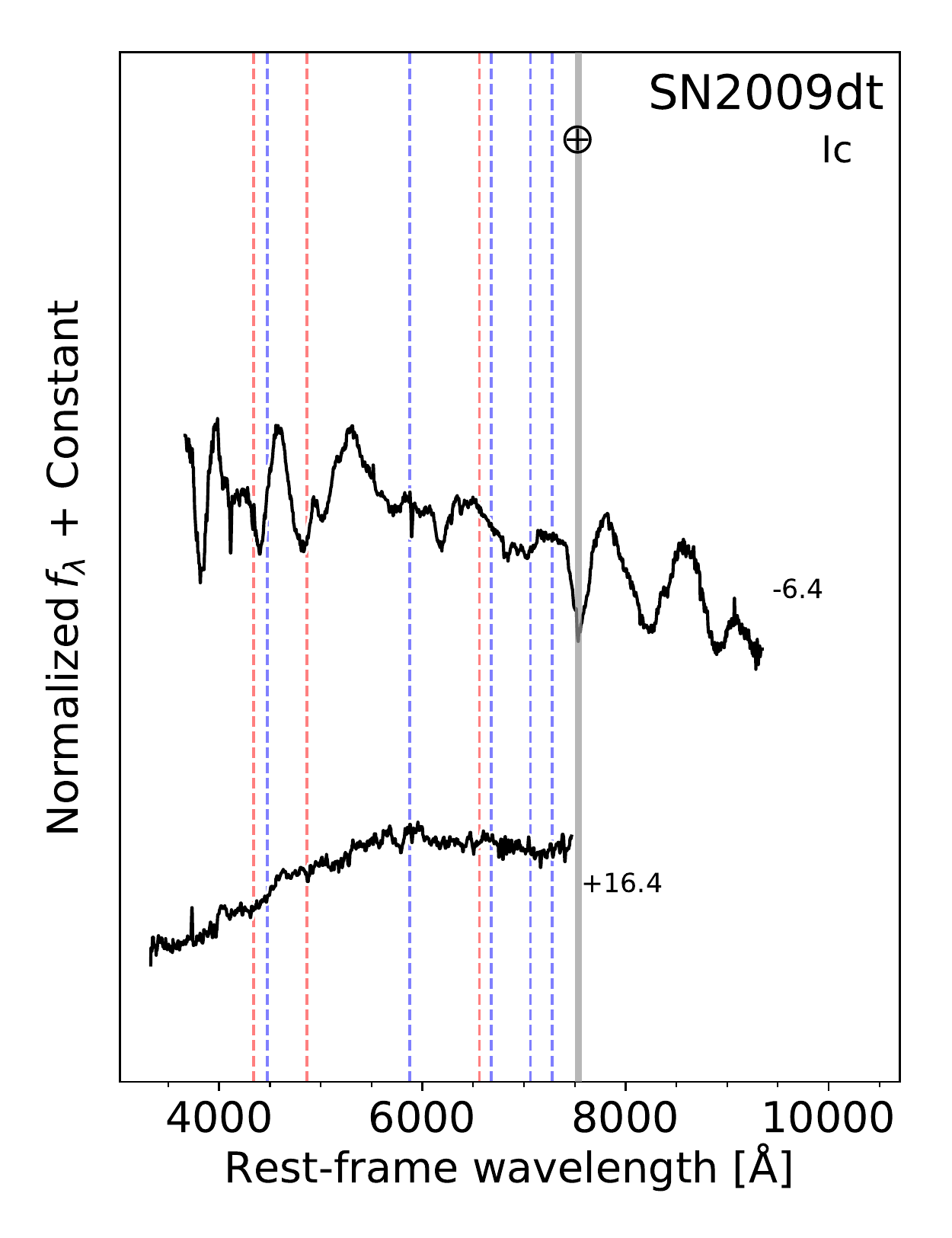}
  \end{array}$
   \caption{Continued.}
   \label{fig:raw_spectra}
\end{figure*}

\begin{figure}[!ht]
  \resizebox{\hsize}{!}
 {\includegraphics[]{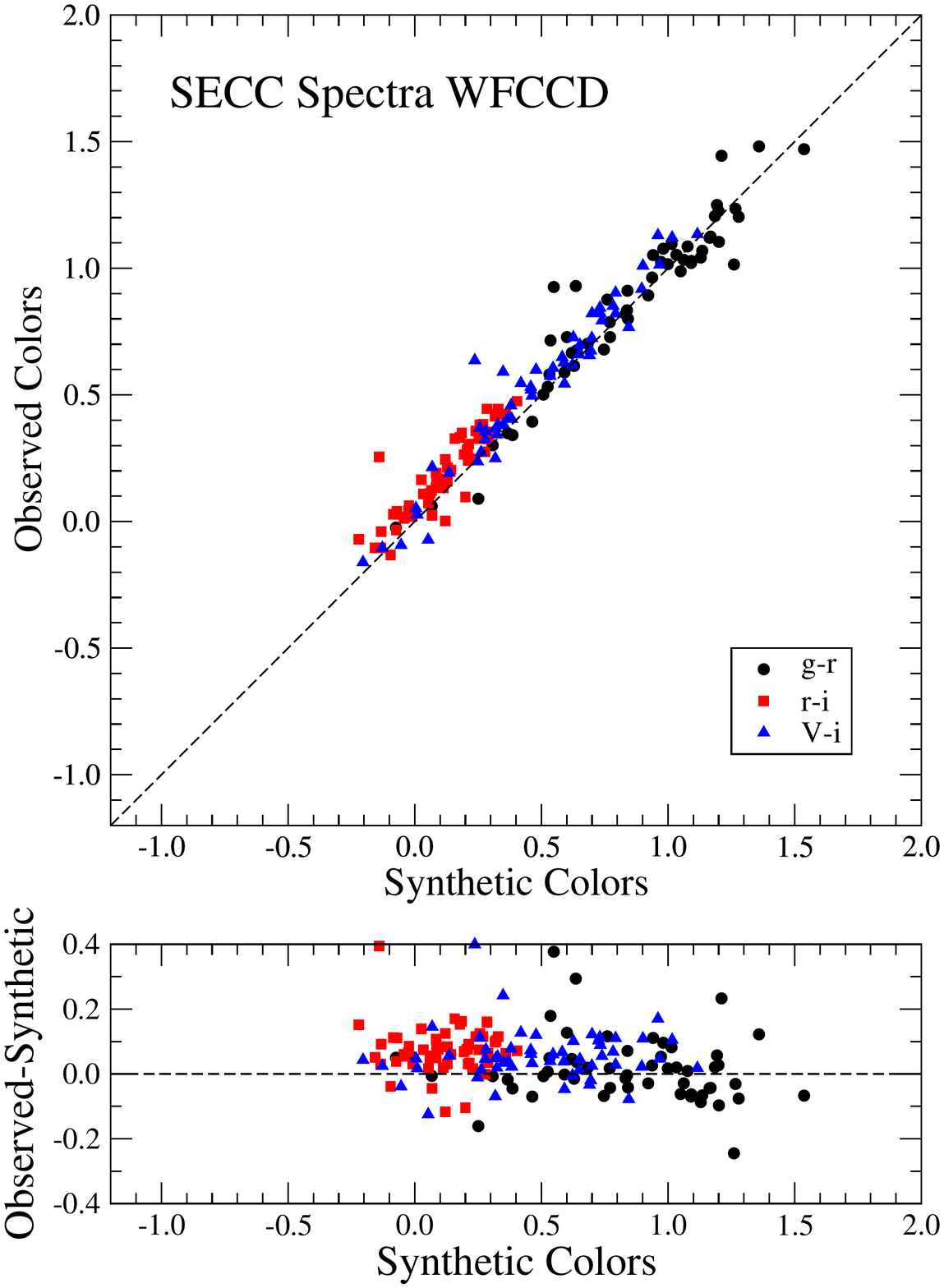}}
  \caption{Quality of the spectral flux calibration. \textit{Top panel:} Comparison between  observed  and synthetic colors ($g-r$) indicated with black dots,  $(r-i$) with red squares, and ($V-i$) with blue triangles. Synthetic colors are computed using WFCCD spectra and the CSP-I system response functions \citep[see][and references therein]{Krisciunas2017}.  
   \textit{Bottom panel:} Residual between  the observed and synthetic colors vs. synthetic colors. The dashed line has a slope of unity.}
   \label{fig:phot_vs_synphot}
\end{figure}

\onecolumn
\begin{appendix}

\section{Journal of spectroscopic observations}

\begin{longtable}{llcccccccc}
\caption{Journal of spectroscopic observations.\label{table2}}\\
\hline
\hline
SN &
UT Date &
JD &
 Phase &
 Telescope &
 Instrument &
Wavelength &
 Resolution &
Exp. time &
 Airmass \\
  &
 &
+2450000 &
  &
  &
  &
 Range [\AA] &
 FWHM [\AA] &
[s] &
  \\
\hline
\endfirsthead
\caption{continued}\\
\hline
SN &
UT Date &
JD &
 Phase &
 Telescope &
 Instrument &
Wavelength &
 Resolution &
Exp. time &
 Airmass \\
  &
 &
+2450000 &
  &
  &
  &
 Range [\AA] &
 FWHM [\AA] &
[s] &
  \\
\hline
\endhead
\hline
\endfoot
\hline
\endlastfoot
2004ew
& 2004 Oct 24.22 & 3302.72 &    26.04 & Clay      & LDSS3      & $3600- 9000$  & 4.2 & 300  & 1.12 \\
& 2004 Nov 26.16 & 3335.66 &    58.28 & Clay      & LDSS3      & $3600- 9000$  & 4.2 & 300  & 1.15 \\
& 2004 Dec 13.14 & 3352.64 &    74.90 & du Pont   & WFCCD      & $3800- 9235$  & 3.0 & 900  & 1.20 \\
& 2004 Dec 19.04 & 3358.54 &    80.67 & Clay      & LDSS3      & $3600- 9000$  & 4.2 & 500  & 1.11 \\
2004ex
& 2004 Oct 24.24 & 3302.74 & $ -$3.88 & Clay      & LDSS3      & $3600- 9000$  & 4.2 & 300  & 1.55 \\
& 2004 Nov 17.18 & 3326.68 &    19.65 & du Pont   & WFCCD      & $3800- 9235$  & 3.0 & 600  & 1.57 \\
& 2004 Nov 26.12 & 3335.62 &    28.43 & Clay      & LDSS3      & $3600- 9000$  & 4.2 & 300  & 1.35 \\
& 2004 Dec 04.09 & 3343.59 &    36.26 & du Pont   & WFCCD      & $3800- 9235$  & 3.0 & 900  & 1.28 \\
& 2004 Dec 09.09 & 3348.59 &    41.18 & du Pont   & WFCCD      & $3800- 9235$  & 3.0 & 900  & 1.35 \\
2004fe
& 2004 Nov 16.06 & 3325.56 &     8.61 & CTIO 1.5m & CS60       & $3200- 9617$  & 5.4 & 1200 & 1.18 \\
& 2004 Nov 17.14 & 3326.64 &     9.68 & du Pont   & WFCCD      & $3800- 9325$  & 3.1 & 600  & 1.32 \\
& 2004 Nov 26.10 & 3335.60 &    18.48 & Clay      & LDSS3   & $3600- 9000$  & 4.2 & 450  & 1.26 \\
& 2004 Dec 13.06 & 3352.56 &    35.14 & du Pont   & WFCCD      & $3800- 9235$  & 3.0 & 900  & 1.29 \\
2004ff
& 2004 Nov 17.28 & 3326.78 &    13.86 & du Pont   & WFCCD      & $3800- 9235$  & 3.0 & 600  & 1.02 \\
& 2004 Nov 26.19 & 3335.69 &    22.57 & Clay      & LDSS3  & $3600- 9000$  & 4.2 & 450  & 1.03 \\
& 2004 Dec 04.22 & 3343.72 &    30.43 & du Pont   & WFCCD      & $3800- 9235$  & 3.0 & 600  & 1.01 \\
2004gq
& 2004 Dec 13.23 & 3352.73 & $-$5.13 & du Pont   & WFCCD      & $3800- 9235$  & 3.0 & 300  & 1.07 \\
& 2004 Dec 19.19 & 3358.69 &     0.78 & Clay      & LDSS3  & $3600- 9000$  & 4.2 & 180  & 1.03 \\
& 2005 Jan 11.16 & 3381.66 &    23.61 & CTIO 1.5m & CS60       & $3200- 9615$  & 5.4 & 900  & 1.09 \\
2004gt
& 2004 Dec 19.30 & 3358.80 & $ -$1.56 & Clay      & LDSS3  & $3600- 9000$  & 4.2 & 180  & 1.58 \\
& 2005 Jan 11.30 & 3381.80 &    21.31 & CTIO 1.5m & CS60       & $3200- 9617$  & 5.4 & 1200 & 1.17 \\
& 2005 Feb 04.32 & 3405.82 &    45.20 & du Pont   & WFCCD      & $3800- 9235$  & 3.0 & 300  & 1.02 \\
& 2005 Feb 08.33 & 3409.83 &    49.19 & du Pont   & WFCCD      & $3800- 9235$  & 3.0 & 300  & 1.02 \\
2004gv
& 2004 Dec 19.09 & 3358.59 & $ -$6.55 & Clay      & LDSS3  & $3600- 9000$  & 4.2 & 300  & 1.18 \\
2005Q
& 2005 Feb 04.06 & 3405.56 & $ -$0.53 & du Pont   & WFCCD      & $3800- 9235$  & 3.0 & 400  & 1.58 \\
& 2005 Feb 12.05 & 3413.56 &     7.29 & du Pont   & WFCCD      & $3800- 9235$  & 3.0 & 400  & 1.76 \\
2005aw
& 2005 Apr 03.33 & 3463.83 &     7.28 & du Pont   & WFCCD      & $3800- 9235$  & 3.0 & 500  & 1.39 \\
& 2005 Apr 07.39 & 3467.89 &    11.30 & du Pont   & WFCCD      & $3800- 9235$  & 3.0 & 600  & 1.16 \\
& 2005 Apr 12.40 & 3472.90 &    16.26 & du Pont   & WFCCD      & $3800- 9235$  & 3.0 & 500  & 1.13 \\
& 2005 Apr 15.39 & 3475.89 &    19.23 & du Pont   & WFCCD      & $3800- 9235$  & 3.0 & 500  & 1.13 \\
& 2005 Apr 19.40 & 3479.90 &    23.20 & du Pont   & WFCCD      & $3800- 9235$  & 3.0 & 500  & 1.11 \\
2005bf
& 2005 Apr 07.07 & 3467.57 & $ -$7.08 & du Pont   & WFCCD      & $3800- 9235$  & 3.0 & 900  & 1.12 \\
& 2005 Apr 12.10 & 3472.60 & $ -$2.15 & du Pont   & WFCCD      & $3800- 9235$  & 3.0 & 900  & 1.12 \\
& 2005 Apr 15.13 & 3475.63 &     0.83 & du Pont   & WFCCD      & $3800- 9235$  & 3.0 & 900  & 1.20 \\
& 2005 Apr 19.01 & 3479.51 &     4.64 & du Pont   & WFCCD      & $3800- 9235$  & 3.0 & 700  & 1.15 \\
& 2005 May 06.97 & 3497.47 &    22.26 & Clay      & LDSS3 & $3838-10000$  & 0.7 & 200  & 1.14 \\
& 2005 May 11.97 & 3502.47 &    27.17 & Clay      & LDSS3  & $6057-10000$  & 1.1 & 900  & 1.12 \\
& 2005 May 31.16 & 3521.67 &    46.01 & du Pont   & WFCCD      & $3800- 8125$  & 3.0 & 1200 & 1.19 \\
& 2005 Jun 01.99 & 3523.49 &    47.79 & du Pont   & WFCCD      & $3800- 8128$  & 3.0 & 1200 & 1.16 \\
& 2005 Jun 06.04 & 3527.54 &    51.77 & du Pont   & WFCCD      & $3800- 8128$  & 3.0 & 600  & 1.48 \\
2005bj
& 2005 Apr 12.35 & 3472.85 &     0.94 & du Pont   & WFCCD      & $3800- 9235$  & $3.0$ & $900$  & 1.47 \\
& 2005 Apr 15.35 & 3475.85 &     3.88 & du Pont   & WFCCD      & $3800- 9235$  & $3.0$ & $800$  & 1.48 \\
2005em
& 2005 Oct 05.35 & 3648.85 & $ -$0.12 & du Pont   & WFCCD      & $3800- 9235$  & $3.0$ & $600$  & 1.21 \\
2006T
& 2006 Feb 13.26 & 3779.76 &     0.15 & ESO NTT   & EMMI       & $3200-10200$  & $2.1$ & $300$  & 1.03 \\
& 2006 Feb 27.35 & 3793.85 &    14.12 & du Pont   & WFCCD      & $3800- 9235$  & $3.0$ & $600$  & 1.72 \\
& 2006 Mar 05.22 & 3799.72 &    19.95 & du Pont   & WFCCD      & $3800- 9235$  & $3.0$ & $600$  & 1.06 \\
& 2006 Mar 08.23 & 3802.73 &    22.93 & du Pont   & WFCCD      & $3800- 9235$  & $3.0$ & $900$  & 1.09 \\
& 2006 Mar 14.21 & 3808.71 &    28.86 & Clay      & LDSS3  & $3785- 9972$  & $1.2$ & $400$  & 1.14 \\
& 2006 Mar 22.19 & 3816.69 &    36.78 & du Pont   & WFCCD      & $3800- 9235$  & $3.0$ & $900$  & 1.09 \\
& 2006 Mar 24.15 & 3818.65 &    38.73 & du Pont   & WFCCD      & $3800- 9235$  & $3.0$ & $900$  & 1.03 \\
& 2006 Apr 02.10 & 3827.60 &    47.61 & du Pont   & WFCCD      & $3800- 9235$  & $3.0$ & $900$  & 1.01 \\
& 2006 Apr 16.24 & 3841.74 &    61.63 & Baade     & IMACS      & $3842- 9692$  & $2.0$ & $600$  & 2.09 \\
& 2006 Apr 23.10 & 3848.60 &    68.43 & du Pont   & WFCCD      & $3800- 9235$  & $3.0$ & $900$  & 1.08 \\
& 2006 Apr 25.05 & 3850.55 &    70.73 & du Pont   & WFCCD      & $3800- 9235$  & $3.0$ & $900$  & 1.02 \\
2006ba
& 2006 Mar 30.15 & 3824.65 &     3.66 & du Pont   & WFCCD      & $3800- 9235$  & $3.0$ & $900$  & 1.13 \\
& 2006 Apr 16.07 & 3841.57 &    20.27 & Baade     & IMACS      & $3842- 9692$  & $2.0$ & $900$  & 1.08 \\
& 2006 Apr 24.04 & 3849.54 &    28.08 & du Pont   & WFCCD      & $3800- 9235$  & $3.0$ & $900$  & 1.07 \\
2006bf
& 2006 Mar 30.31 & 3824.81 &     7.18 & du Pont   & WFCCD      & $3800- 9235$  & $3.0$ & $900$  & 1.54 \\
& 2006 Apr 02.32 & 3827.82 &    10.12 & du Pont   & WFCCD      & $3800- 9235$  & $3.0$ & $1200$ & 1.67 \\
& 2006 Apr 22.25 & 3847.75 &    29.58 & du Pont   & WFCCD      & $3800- 9235$  & $3.0$ & $1000$ & 1.53 \\
& 2006 Apr 23.23 & 3848.73 &    30.54 & du Pont   & WFCCD      & $3800- 9235$  & $3.0$ & $1000$ & 1.47 \\
& 2006 Apr 25.28 & 3850.78 &    32.55 & du Pont   & WFCCD      & $3800- 9235$  & $3.0$ & $900$  & 2.04 \\
2006ep
& 2006 Sep 26.20 & 4004.70 &    19.00 & du Pont   & WFCCD      & $3800- 9235$  & $3.0$ & $1200$ & 1.74 \\
& 2006 Oct 10.19 & 4018.69 &    32.79 & du Pont   & WFCCD      & $3800- 9235$  & $3.0$ & $1200$ & 1.74 \\
& 2006 Oct 11.18 & 4019.68 &    33.76 & du Pont   & WFCCD      & $3800- 9235$  & $3.0$ & $1200$ & 1.73 \\
2006fo
& 2006 Sep 26.28 & 4004.78 &     1.58 & du Pont   & WFCCD      & $3800- 9235$  & $3.0$ & $1200$ & 1.16 \\
& 2006 Sep 28.26 & 4006.76 &     3.52 & du Pont   & WFCCD      & $3800- 9235$  & $3.0$ & $1200$ & 1.18 \\
& 2006 Oct 13.32 & 4021.82 &    18.27 & du Pont   & WFCCD      & $3800- 9235$  & $3.0$ & $1200$ & 1.28 \\
2006ir
& 2006 Oct 10.10 & 4018.60 &    19.14 & du Pont   & WFCCD      & $3800- 9235$  & $3.0$ & $600$  & 1.25 \\
& 2006 Nov 03.14 & 4042.64 &    42.71 & ESO NTT   & EMMI       & $4012-10200$  & $3.3$ & $300$  & 1.53 \\
2006lc
& 2006 Nov 03.08 & 4042.58 &     1.23 & ESO NTT   & EMMI       & $3200-10200$  & $2.1$ & $300$  & 1.20 \\
2007C
& 2007 Jan 13.34 & 4113.84 & $ -$1.40 & du Pont   & B\&C       & $3932-10142$  & $3.0$ & $300$  & 1.29 \\
& 2007 Jan 18.28 & 4118.78 &     3.51 & du Pont   & B\&C       & $3402- 9614$  & $3.0$ & $900$  & 1.63 \\
& 2007 Jan 31.36 & 4131.86 &    16.52 & ESO NTT   & EMMI       & $3200- 5300$  & $2.1$ & $400$  & 1.10 \\
& 2007 Feb 12.37 & 4143.87 &    28.46 & du Pont   & WFCCD      & $3800- 9235$  & $3.0$ & $400$  & 1.09 \\
& 2007 Feb 19.38 & 4150.88 &    35.43 & du Pont   & WFCCD      & $3800- 9235$  & $3.0$ & $400$  & 1.13 \\
& 2007 Feb 25.35 & 4156.85 &    41.37 & du Pont   & WFCCD      & $3800- 9235$  & $3.0$ & $400$  & 1.10 \\
& 2007 Mar 04.27 & 4163.77 &    48.25 & ESO NTT   & EMMI       & $4000-10200$  & $3.3$ & $300$  & 1.09 \\
& 2007 Mar 19.22 & 4178.72 &    63.12 & du Pont   & B\&C       & $3419- 9626$  & $3.0$ & $900$  & 1.11 \\
& 2007 Apr 17.30 & 4207.80 &    92.04 & du Pont   & WFCCD      & $3800- 9235$  & $3.0$ & $600$  & 1.51 \\
2007Y
& 2007 Feb 19.04 & 4150.54 & $-$12.39 & du Pont   & WFCCD      & $3800- 9235$  & $3.0$ & $600$  & 1.31 \\
& 2007 Feb 25.02 & 4156.52 & $ -$6.44 & du Pont   & WFCCD      & $3800- 9235$  & $3.0$ & $600$  & 1.28 \\
& 2007 Mar 04.05 & 4163.55 &     0.56 & ESO NTT   & EMMI       & $3200-10200$  & $2.1$ & $200$  & 1.76 \\
& 2007 Mar 11.00 & 4170.50 &     7.47 & Baade     & IMACS      & $3789-10879$  & $2.0$ & $600$  & 1.42 \\
& 2007 Mar 14.01 & 4173.51 &    10.47 & du Pont   & B\&C       & $3429- 9648$  & $3.0$ & $600$  & 1.62 \\
& 2007 Mar 19.00 & 4178.50 &    15.44 & du Pont   & B\&C       & $3419- 9631$  & $3.0$ & $600$  & 1.60 \\
& 2007 Mar 25.99 & 4185.49 &    22.39 & du Pont   & B\&C       & $3400- 9606$  & $3.0$ & $600$  & 1.73 \\
& 2007 Apr 12.98 & 4203.48 &    40.30 & du Pont   & WFCCD      & $3800- 9235$  & $3.0$ & $800$  & 2.59 \\
& 2007 Oct 21.32 & 4394.82 &   230.76 & Clay      & LDSS3  & $4097- 9979$  & $0.7$ & $900$  & 1.11 \\
& 2007 Oct 23.30 & 4396.80 &   232.73 & Clay      & LDSS3  & $4033- 9988$  & $0.7$ & $1200$ & 1.06 \\
& 2007 Nov 17.28 & 4421.78 &   257.60 & Baade     & IMACS      & $3809-10675$  & $2.0$ & $1800$ & 1.21 \\
& 2007 Dec 01.21 & 4435.71 &   271.46 & ESO 3.6m  & EFOSC      & $3300- 9260$  & $2.8$ & $1200$ & 1.09 \\
2007ag
& 2007 Mar 11.17 & 4170.67 &     7.23 & Baade     & IMACS      & $3789-10887$  & $2.0$ & $600$  & 1.59 \\
& 2007 Mar 14.13 & 4173.63 &    10.13 & du Pont   & B\&C       & $3429- 9642$  & $3.0$ & $1200$ & 1.15 \\
2007hn
& 2007 Sep 18.20 & 4361.70 &     9.80 & du Pont   & B\&C       & $3408- 9614$  & $3.0$ & $1200$ & 1.43 \\
& 2007 Oct 04.03 & 4377.53 &    25.21 & ESO 3.6m  & EFOSC      & $3300- 9260$  & $2.8$ & $1200$ & 1.11 \\
& 2007 Oct 05.05 & 4378.55 &    26.20 & Clay      & LDSS3  & $4452- 9303$  & $2.0$ & $1200$ & 1.10 \\
& 2007 Nov 05.01 & 4409.51 &    56.34 & ESO 3.6m  & EFOSC      & $3300- 9260$  & $2.7$ & $1200$ & 1.16 \\
2007kj
& 2007 Oct 03.15 & 4376.65 & $ -$4.18 & ESO 3.6m  & EFOSC      & $3300- 9260$  & $2.8$ & $600$  & 1.36 \\
& 2007 Oct 05.12 & 4378.63 & $ -$2.24 & Clay      & LDSS3  & $4452- 9303$  & $2.0$ & $600$  & 1.39 \\
& 2007 Oct 16.18 & 4389.68 &     8.61 & du Pont   & B\&C       & $3412- 9624$  & $3.0$ & $900$  & 1.40 \\
& 2007 Oct 21.13 & 4394.63 &    13.48 & du Pont   & B\&C       & $3352- 9563$  & $3.0$ & $1200$ & 1.35 \\
& 2007 Nov 17.10 & 4421.60 &    39.98 & Baade     & IMACS      & $3809-10675$  & $2.0$ & $600$  & 1.46 \\
2007rz
& 2007 Dec 10.26 & 4444.76 &     7.35 & du Pont   & WFCCD      & $3800- 9235$  & $3.0$ & $700$  & 1.48 \\
& 2008 Jan 04.12 & 4469.62 &    31.89 & ESO 3.6m  & EFOSC      & $3300- 9260$  & $2.8$ & $600$  & 1.26 \\
2008aq
& 2008 Feb 29.31 & 4525.81 & $ -$5.30 & Baade     & IMACS      & $3743-10721$  & $2.0$ & $600$  & 1.06 \\
& 2008 Mar 14.32 & 4539.82 &     8.60 & ESO 3.6m  & EFOSC      & $3300- 9240$  & $2.7$ & $600$  & 1.12 \\
& 2008 Mar 19.31 & 4544.81 &    13.55 & Clay      & MagE       & $3101- 9344$  & $0.3$ & $1200$ & 1.13 \\
& 2008 Mar 20.32 & 4545.82 &    14.55 & Clay      & LDSS3  & $3629- 9437$  & $2.0$ & $600$  & 1.19 \\
& 2008 Mar 28.33 & 4553.83 &    22.50 & ESO NTT   & EMMI       & $3200-10200$  & $2.1$ & $300$  & 1.35 \\
& 2008 Mar 31.29 & 4556.79 &    25.43 & du Pont   & WFCCD      & $3800- 9235$  & $3.0$ & $600$  & 1.18 \\
& 2008 Apr 12.27 & 4568.77 &    37.33 & du Pont   & WFCCD      & $3800- 9235$  & $3.0$ & $600$  & 1.26 \\
& 2008 Apr 26.22 & 4582.72 &    51.16 & Baade     & IMACS      & $4246-10374$  & $2.0$ & $600$  & 1.17 \\
& 2008 Apr 29.19 & 4585.69 &    54.11 & du Pont   & WFCCD      & $3800- 9235$  & $3.0$ & $600$  & 1.13 \\
& 2008 May 11.15 & 4597.65 &    65.98 & du Pont   & WFCCD      & $3800- 9235$  & $3.0$ & $600$  & 1.12 \\
& 2008 May 22.15 & 4608.65 &    76.89 & Baade     & IMACS      & $4338-10868$  & $2.0$ & $600$  & 1.19 \\
& 2009 Jan 10.28 & 4841.78 &   308.17 & Gemini-S  & GMOS       & $3900- 8118$  & $1.4$ & $1800$ & 1.53 \\
2008gc
& 2008 Oct 14.32 & 4753.82 &     9.11 & ESO NTT   & EFOSC      & $3401- 9210$  & $2.8$ & $1200$ & 1.26 \\
& 2008 Oct 15.15 & 4754.65 &     9.90 & ESO NTT   & EFOSC      & $3351- 9210$  & $2.7$ & $1200$ & 1.17 \\
& 2008 Oct 20.22 & 4759.72 &    14.74 & du Pont   & WFCCD      & $3800- 9235$  & $3.0$ & $900$  & 1.10 \\
& 2008 Oct 28.30 & 4767.80 &    22.43 & du Pont   & WFCCD      & $3800- 9235$  & $3.0$ & $1200$ & 1.25 \\
& 2008 Nov 04.07 & 4774.57 &    28.89 & Clay      & LDSS3  & $3636- 9452$  & $2.0$ & $1000$ & 1.23 \\
& 2009 Jan 30.04 & 4861.54 &   117.28 & Gemini-S  & GMOS       & $3902- 8118$  & $1.4$ & $2400$ & 1.32 \\
2008hh
& 2008 Nov 23.12 & 4793.62 &     2.08 & du Pont   & WFCCD      & $3800- 9235$  & $3.0$ & $600$  & 1.36 \\
& 2008 Dec 08.10 & 4808.60 &    16.78 & Baade     & IMACS      & $4023-10719$  & $2.0$ & $900$  & 1.41 \\
& 2008 Dec 23.04 & 4823.54 &    31.43 & du Pont   & WFCCD      & $3800- 9235$  & $3.0$ & $1200$ & 1.35 \\
2009K
& 2009 Jan 17.09 & 4848.59 & $-$18.42 & Clay      & LDSS3  & $3629- 9440$  & $2.0$ & $500$  & 1.15 \\
& 2009 Jan 22.08 & 4853.58 & $-$13.48 & Clay      & LDSS3  & $3620- 9428$  & $2.0$ & $300$  & 1.16 \\
& 2009 Feb 08.13 & 4870.63 &     3.37 & Clay      & LDSS3  & $3716- 9438$  & $2.0$ & $500$  & 1.59 \\
& 2009 Feb 11.07 & 4873.57 &     6.28 & Baade     & IMACS      & $4032-10122$  & $2.0$ & $500$  & 1.27 \\
& 2009 Feb 25.08 & 4887.58 &    20.13 & du Pont   & WFCCD      & $3800- 9235$  & $3.0$ & $500$  & 1.61 \\
& 2009 Mar 16.00 & 4906.50 &    38.83 & ESO NTT   & EFOSC      & $3300- 9210$  & $2.7$ & $500$  & 1.57 \\
& 2009 Mar 28.02 & 4918.53 &    50.71 & du Pont   & WFCCD      & $3800- 9235$  & $3.0$ & $500$  & 1.93 \\
& 2009 Nov 17.33 & 5152.84 &   282.31 & Gemini-S  & GMOS       & $4099- 7999$  & $1.4$ &$1350$  & 1.50 \\
2009Z
& 2009 Feb 08.35 & 4870.85 & $ -$5.95 & Clay      & LDSS3  & $3716- 9439$  & $2.0$ & $900$  & 1.20 \\
& 2009 Feb 11.35 & 4873.85 & $ -$3.01 & Baade     & IMACS      & $4032-10120$  & $2.0$ & $800$  & 1.17 \\
& 2009 Feb 15.37 & 4877.87 &     0.91 & Baade     & IMACS      & $4021-10121$  & $2.0$ & $600$  & 1.13 \\
& 2009 Feb 24.35 & 4886.85 &     9.67 & du Pont   & WFCCD      & $3800- 9235$  & $3.0$ & $700$  & 1.13 \\
& 2009 Feb 25.31 & 4887.81 &    10.61 & du Pont   & WFCCD      & $3800- 9235$  & $3.0$ & $900$  & 1.18 \\
& 2009 Feb 26.37 & 4888.87 &    11.65 & du Pont   & WFCCD      & $3800- 9235$  & $3.0$ & $900$  & 1.14 \\
& 2009 Mar 15.36 & 4905.86 &    28.22 & Baade     & IMACS      & $4026-10114$  & $2.0$ & $900$  & 1.20 \\
& 2009 Mar 28.37 & 4918.87 &    40.92 & du Pont   & WFCCD      & $3800- 9235$  & $3.0$ & $900$  & 1.41 \\
2009bb
& 2009 Mar 28.19 & 4918.69 & $ -$1.35 & du Pont   & WFCCD      & $3800- 9235$  & $3.0$ & $700$  & 1.09 \\
& 2009 Mar 29.10 & 4919.60 & $ -$0.45 & du Pont   & WFCCD      & $3800- 9235$  & $3.0$ & $900$  & 1.03 \\
& 2009 Apr 03.14 & 4924.64 &     4.55 & du Pont   & WFCCD      & $3800- 9235$  & $3.0$ & $700$  & 1.04 \\
& 2009 Apr 07.00 & 4928.50 &     8.37 & Baade     & IMACS      & $4057-10139$  & $2.0$ & $400$  & 1.16 \\
& 2009 Apr 15.31 & 4936.81 &    16.60 & Gemini-S  & GMOS       & $3904- 8122$  & $1.4$ & $900$  & 1.03 \\
& 2009 Apr 17.10 & 4938.60 &    18.37 & Clay      & LDSS3  & $3714- 9435$  & $2.0$ & $600$  & 1.03 \\
& 2009 Apr 18.08 & 4939.58 &    19.34 & du Pont   & B\&C       & $3450- 9657$  & $3.0$ & $700$  & 1.02 \\
& 2009 Apr 22.08 & 4943.58 &    23.30 & du Pont   & B\&C       & $3360- 9567$  & $3.0$ & $900$  & 1.03 \\
& 2009 Apr 23.11 & 4944.61 &    24.32 & du Pont   & B\&C       & $3362- 9565$  & $3.0$ & $1000$ & 1.07 \\
& 2009 Apr 26.31 & 4947.81 &    27.49 & Gemini-S  & GMOS       & $3904- 8122$  & $1.4$ & $900$  & 1.03 \\
& 2009 Apr 30.13 & 4951.64 &    31.28 & Clay      & LDSS3  & $3702- 9443$  & $2.0$ & $600$  & 1.18 \\
& 2009 May 01.08 & 4952.58 &    32.21 & Clay      & LDSS3  & $3721- 9445$  & $2.0$ & $600$  & 1.05 \\
& 2009 May 14.03 & 4965.53 &    45.04 & Baade     & IMACS      & $3974-10136$  & $2.0$ & $600$  & 1.04 \\
& 2009 May 22.96 & 4974.46 &    53.88 & du Pont   & B\&C       & $3360- 9565$  & $3.0$ & $900$  & 1.02 \\
& 2009 May 31.05 & 4982.55 &    61.89 & du Pont   & B\&C       & $3394- 9591$  & $3.0$ & $1200$ & 1.17 \\
& 2010 Jan 09.21 & 5205.71 &   282.86 & Clay      & LDSS3  & $3632- 9441$  & $2.0$ & $1800$ & 1.33 \\
& 2010 Jan 10.29 & 5206.79 &   283.93 & Gemini-S  & GMOS       & $4197- 8424$  & $1.4$ & $4000$ & 1.02 \\
& 2010 Feb 03.09 & 5230.59 &   307.50 & Baade     & IMACS      & $3975-10051$  & $2.0$ & $1800$ & 1.75 \\
2009ca
& 2009 Apr 07.42 & 4928.92 &     2.51 & Baade     & IMACS      & $4057-10139$  & $2.0$ & $900$  & 1.31 \\
& 2009 Apr 17.35 & 4938.85 &    11.57 & Clay      & LDSS3  & $3714- 9435$  & $2.0$ & $700$  & 1.62 \\
& 2009 Apr 22.37 & 4943.87 &    16.15 & du Pont   & B\&C       & $3360- 9567$  & $3.0$ & $900$  & 1.36 \\
2009dp
& 2009 Apr 30.33 & 4951.83 &     1.98 & Clay      & LDSS3  & $3702- 9445$  & $2.0$ & $600$  & 1.28 \\
& 2009 May 01.37 & 4952.87 &     2.99 & Clay      & LDSS3  & $3721- 9445$  & $2.0$ & $600$  & 1.11 \\
& 2009 Jun 02.28 & 4984.78 &    34.18 & Clay      & MagE       & $3181- 9111$  & $0.2$ & $900$  & 1.12 \\
2009dq
& 2009 Apr 30.10 & 4951.60 & $ -$8.85 & Clay      & LDSS3  & $3702- 9422$  & $2.0$ & $400$  & 1.36 \\
& 2009 May 23.05 & 4974.55 &    13.98 & du Pont   & B\&C       & $3360- 9563$  & $3.0$ & $600$  & 1.37 \\
2009dt
& 2009 Apr 30.36 & 4951.86 & $ -$6.38 & Clay      & LDSS3  & $3702- 9445$  & $2.0$ & $600$  & 1.44 \\
& 2009 May 23.39 & 4974.89 &    16.41 & du Pont   & B\&C       & $3360- 9565$  & $3.0$ & $900$  & 1.06 \\
\end{longtable}

\end{appendix}
\end{document}